\documentclass[twocolumn,nofootinbib,superscriptaddress,floatfix,usenames,dvipsnames]{revtex4-2}\pdfoutput=1

\usepackage{amsmath}
\usepackage{amsfonts}
\usepackage[retainorgcmds]{IEEEtrantools}
\usepackage[colorlinks=true, linktoc=page, citecolor=blue, urlcolor=blue]{hyperref}
\usepackage{scalerel}
\usepackage{tikz-feynman}
\usepackage{float}
\usepackage{colortbl}
\usepackage{multirow}

\newcommand{\quotes}[1]{``#1''}
\newcommand{\N}{\mathcal{N}}
\newcommand{\CC}{\mathcal{C}}
\newcommand{\BB}{\mathcal{B}}
\newcommand{\II}{\mathcal{I}}
\newcommand{\JJ}{\mathcal{J}}
\newcommand{\NN}{\mathfrak{N}}
\newcommand{\MM}{\mathcal{M}}
\newcommand{\rD}{\mathring{\Delta}}
\newcommand{\gym}{g_{\text{\scalebox{.6}{YM}}}}
\newcommand{\tn}{\textnormal}

\newcommand{\tr}{{\text{tr}}}
\def\aacute#1{\text{\fontfamily{cmr}\selectfont \textit{\H{#1}}}}

\hypersetup{pdfstartview={XYZ null null 0.75}}
\let\oldpageref\pageref\renewcommand{\pageref}{\oldpageref*}

\makeatletter\renewcommand{\p@subsection}{{\thesection.}}\makeatother
\makeatletter\renewcommand{\p@subsubsection}{{\thesection.\thesubsection.}}\makeatother

\begin{document}

\title{\texorpdfstring{\color{Blue}\textbf{Heavy holographic correlators in defect conformal field theories}}{Heavy holographic correlators in defect conformal field theories}}
\author{\textbf{Georgios Linardopoulos}}\email{george.linardopoulos@simis.cn}\affiliation{Shanghai Institute for Mathematics and Interdisciplinary Sciences (SIMIS),\\ Shanghai 200433, China.}
\author{\textbf{Chanyong Park}}\email{cyong21@gist.ac.kr}\affiliation{Department of Physics and Photon Science, Gwangju Institute of Science \\ and Technology, Gwangju 61005, Korea}

\begin{abstract}
\noindent We study holographic defect conformal field theories which are dual to probe branes with bottom-up methods. First we determine the embedding of codimension-1 interface branes in AdS space. Then we compute defect one and two-point functions of heavy scalar operators at strong coupling. We use geodesic approximations to compute reflected two-point functions in two ways, as well as ambient and defect-channel two-point functions. In appropriate limits, our results agree with the operator product expansion (OPE) and the boundary operator expansion (BOE).
\end{abstract}

\maketitle
\section{Introduction}
\noindent Understanding strongly interacting systems is one of the greatest challenges in theoretical physics. As traditional Quantum Field Theory (QFT) lacks a systematic framework for exploring nonperturbative phenomena, an effective tool is provided by the AdS/CFT correspondence and holography \cite{Maldacena97, GubserKlebanovPolyakov98, Witten98a, Witten98b}. The holographic proposal claims that $(d+1)$ dimensional (weakly coupled, super) gravity/string theories in AdS spaces are in a one-to-one mapping with $d$-dimensional (strongly interacting, maximally supersymmetric) conformal field theories (CFTs). As such, holography paves the way for investigating strongly interacting systems. Holographic QFTs with fewer symmetries and supersymmetries have also been identified, e.g.\ holographic boundary and defect CFTs (BCFTs and dCFTs), where planar boundaries and defects of various codimensionalities break translation invariance \cite{KarchRandall01a, KarchRandall01b}. \\
\indent The computation of strongly coupled correlation functions with the AdS/CFT correspondence may proceed with the top-down and the bottom-up approach.\footnote{See e.g.\ \cite{HarveyJensenUzu25} for a recent description of the two approaches.} In the top-down approach, the detailed gravity/string theory duals of QFTs are known exactly. Leading order correlation functions can be computed with Witten diagrams in the supergravity approximation \cite{FreedmanMathurMatusisRastelli98a, ChalmersNastaseSchalmSiebelink98, LeeMinwallaRangamaniSeiberg98, ArutyunovFrolov99b, Lee99}, whereas for the next-to-leading orders beyond supergravity, semiclassical string methods are available. There is a distinction between scaling dimensions (spectrum) and two-point functions \cite{GubserKlebanovPolyakov02, JanikSurowkaWereszczynski10, Buchbinder10, BuchbinderTseytlin10a, Tseytlin03a}, whose computation generally differs from that of three and higher-point correlation functions \cite{BerensteinCorradoFischlerMaldacena99, Zarembo10c, CostaMonteiroSantosZoakos10}. In integrable models of AdS/CFT \cite{MinahanZarembo03, BeisertKristjansenStaudacher04, Beisert03}, all-loop nonperturbative methods for both the spectrum \cite{GromovKazakovLeurentVolin13} and correlation functions \cite{BassoKomatsuVieira15} have been developed.\footnote{See also \cite{RigatosZhou22} for an interesting new connection between holography and integrability which is worth exploring further.} \\
\indent Top-down models of boundary/defect CFTs which have concrete holographic realizations with probe branes, go by the name of AdS/defect CFT correspondence (AdS/{\color{Red}d}CFT for short). For such models, supergravity/Witten diagram computations have been found to agree with perturbative QFT calculations (in appropriate scaling limits). Known cases include the (codimension-1) probe brane systems D3-D5 \cite{NagasakiTanidaYamaguchi11, NagasakiYamaguchi12, deLeeuwKristjansenZarembo15}, D3-D7 \cite{KristjansenSemenoffYoung12b}, and D2-D4 \cite{KristjansenVuZarembo21}, the (codimension-3) D1-D3 system \cite{KristjansenZarembo23} and the (codimension-2) D3-D3 \cite{DrukkerGomisMatsuura08} and D3-D5 \cite{GeorgiouLinardopoulosZoakos25} probe brane systems. Semiclassical string methods for these systems were introduced in \cite{GeorgiouLinardopoulosZoakos23} (see \cite{Linardopoulos25a} for a recent review). Localization techniques for supersymmetric probe brane systems have also been developed \cite{RobinsonUhlemann17, Robinson17, Wang20a, KomatsuWang20, BeccariaCaboBizet23}. \\
\indent On the other hand, the bottom-up approach begins with some generic AdS type of background and successively adds to it new ingredients in order to approximate some desired holographic dual QFT. This way, physical quantities of QFTs such as the $q \bar{q}$ potential \cite{Maldacena98, Park09, Park11}, entanglement entropy \cite{RyuTakayanagi06a, RyuTakayanagi06b, Park15c, KimPark16b, NarayananParkZhang18, Park15b, Park18a} and correlation functions \cite{SusskindWitten98, Solodukhin98b, BalasubramanianRoss99, LoukoMarolfRoss00, KrausOoguriShenker03, KimPalPark23, Park24, ParkLee24} are realized by geometric objects on the dual gravity side. For two heavy scalar local operators located on the boundary of AdS, their two-point correlation function can be computed at strong coupling from the length of the geodesic curve which connects them in the bulk of AdS \cite{SusskindWitten98, Solodukhin98b, BalasubramanianRoss99, LoukoMarolfRoss00, KrausOoguriShenker03}:
\begin{IEEEeqnarray}{c}
\langle \hat{O}(x)\hat{O}(y) \rangle \simeq e^{-\Delta\cdot L(x; y)/R}, \label{TwoPointFunctionCFT1}
\end{IEEEeqnarray}
where $\Delta$ is the scaling dimension of the scalar operator $\hat{O}(x)$, $L(x; y)$ is the geodesic distance of the boundary points $x, y$, and $R$ is the AdS length. With this holographic proposal, the known expressions of conformal two and three-point functions were obtained in \cite{KimPalPark23}. \\
\indent Bottom-up methods typically ignore CFT matter fields and their dual components in the bulk by focusing on the classical gravity limit of the AdS/CFT correspondence. The latter is obtained by allowing the number of colors $N$ and 't Hooft coupling $\lambda = \gym^2 N$ to become infinite. The new ingredient we add in this work is a bulk codimension-1 probe brane. As mentioned above, probe branes are holographically dual to BCFTs and dCFTs. The main goal of this paper is the computation of holographic correlation functions at strong coupling with the bottom-up approach. Because bottom-up holography takes into account only salient features of the models under study, details which are needed for the identification of string states and their dual QFT operators are by and large glossed over. Even so, bottom-up methods provide useful information about the structure of holographic correlation functions at strong coupling, in the limit where the involved operators become heavy as we will see. \\
\indent Gauge invariant operators in the large-$N$ limit of holographic CFTs like $\N = 4$, $SU(N)$ super Yang-Mills (SYM) theory are classified according to their scaling dimensions. See table \ref{Table:OperatorTypes} below.\footnote{A nice recent reference where more details can be found is \cite{deMelloKochKimMahu24}. See also the classification of \cite{AnempodistovHolguinKazakovMurali25}.} Light operators have small and finite scaling dimensions, while the dimensions of heavier states tend to infinity with increasing powers of $N$. Single-trace operators which diagonalize the planar ($N \rightarrow \infty$), integrable part of the dilatation operator of $\N = 4$ SYM cannot be heavier than $\sqrt{N}$. Heavier operators which scale like $N$ include giant gravitons, which are probe finite-size compact D-branes, holographically dual to determinant operators. Huge states can also be constructed, e.g.\ by putting together stacks of D-branes which backreact with the background geometry and are believed to be relevant to black holes.\\
\renewcommand{\arraystretch}{1.5}\setlength{\tabcolsep}{5pt}
\begin{table}[H]\begin{center}\begin{tabular}{|l||l|c|}
\hline
& \hspace{.2cm} Scaling & \multirow{2}{*}{Examples} \\[-6pt] & dimensions & \\ \hline\hline
Light operators & \hspace{.1cm} $\Delta \sim 1 $ & CPOs \\ \hline
Medium operators & \hspace{.1cm} $\Delta \sim N^{1/4}$ & Konishi operator \\ \hline
Heavy operators & \hspace{.1cm} $\Delta \sim N^{1/2}$ & giant magnons \\ \arrayrulecolor{Red}\hline\arrayrulecolor{black}
Heavier operators & \hspace{.1cm} $\Delta \sim N $ & giant gravitons \\ \hline
Huge operators & \hspace{.1cm} $\Delta \sim N^2$ & backreacting operators \\ \hline
\end{tabular}\caption{Types of gauge invariant operators in $\N = 4$ SYM. \label{Table:OperatorTypes}}\end{center}\end{table}
\vspace{-.5cm}\indent The computation of heavy correlation functions is currently receiving increased attention, especially in the case of huge backreacting operators \cite{GrinbergMaldacena20, Holguin23, AbajianAprileMyersVieira23a, AbajianAprileMyersVieira23b, KazakovMuraliVieira24}. Recent studies of the heavy OPE can be found in \cite{PolandRogelberg25, PolandRogelberg26b}. Perturbative calculations of heavy correlation functions become more and more involved as the number of Wick contractions increases and usually stop after a few loop orders. Strong coupling calculations are essentially limited to holographic models. Top-down methods generally start with a saddle point approximation which makes the path integral localize on classical solutions, to leading order in the coupling ($\lambda\rightarrow\infty$). In the heavy operator limit ($\Delta\rightarrow\infty$), the leading order correlator further localizes on geodesic curves. By contrast, the bottom-up approach directly employs the geodesic approximation recipe \eqref{TwoPointFunctionCFT1} and omits the preliminary steps which are included in the top-down algorithm. This simplifies and speeds up the computation of correlation functions significantly, however several details of the correlators (such as e.g.\ the precise form of the operators) are no longer available. \\
\indent Our paper is organized as follows. In section \ref{Section:TopDownApproach} below we revisit the top-down computation of one-point functions of chiral primary operators (CPOs) with the AdS/{\color{Red}d}CFT correspondence. In section \ref{Section:BottomUpApproach} we introduce the bottom-up approach and show that it reproduces the embedding of codimension-1 probe branes in AdS spacetime. We then compute correlation functions of heavy scalar operators in strongly coupled codimension-1 dCFT with the method of geodesic approximations. In subsection \ref{SubSection:OnePointFunctions} we compute one-point functions, and in subsection \ref{SubSection:TwoPointFunctions} two-point functions. In \S\ref{SubSubSection:TwoPointReflected} we compute reflected two-point functions in two ways, in \S\ref{SubSubSection:TwoPointAmbient} we compute defect CFT two-point functions in the ambient channel and in \S\ref{SubSubSection:TwoPointDefect} defect CFT two-point functions in the defect channel. Section \ref{Section:Conclusions} includes our concluding remarks.
\section{Top-down approach \label{Section:TopDownApproach}}
\noindent We consider holographic models of defect CFTs whose string/gravity theory duals include a probe brane. Top-down (AdS/{\color{Red}d}CFT) descriptions of these models are obtained by adding probe branes on the dual string theory sides of integrable holographic (supersymmetric) gauge theories and CFTs, such as planar $\N = 4$ super Yang-Mills (SYM) theory in 4d and ABJM theory in 3d. The probe brane induces a defect on the dual CFT side which becomes a dCFT. Supersymmetry and/or integrability may be broken on either side of the emerging duality. \\
\indent For example, take two copies of $\N = 4$ SYM theory, with gauge groups $SU(N)$ and $SU(N-k)$, separated by a (planar) 3d codimension-1 defect at $x_3 = 0$.\footnote{The AdS$_5$ Poincaré coordinates are denoted by $(x_0,x_1,x_2,x_3,z)$, while the coordinates on S$^5$ are $(\psi,\theta,\chi,\vartheta,\varrho)$. See \eqref{AdS5xS5metric} below and \eqref{SphereParametrization1}--\eqref{S5metric} in appendix \ref{AppendixA}. This implies that the boundary of AdS$_5$ is located at $z = 0$. Note also the notation $x = (x_0,x_1,x_2,x_3)$ and $\textbf{x} = (x_0,x_1,x_2)$.} Such a system can be realized in string theory by a probe D5 or D7-brane which is anchored to the boundary of AdS$_5$ and wraps an internal space with $k$ units of Abelian or non-Abelian flux through it \cite{NagasakiTanidaYamaguchi11, NagasakiYamaguchi12, KristjansenSemenoffYoung12b}.
\subsection{Weak coupling}
\noindent AdS/{\color{Red}d}CFT correlation functions can be computed by means of \quotes{interfaces}. Interfaces are described by classical (\quotes{fuzzy funnel} or \quotes{domain wall}) solutions of the gauge theory equations of motion which share the symmetries of the probe brane \cite{ConstableMyersTafjord99, ConstableMyersTafjord01a}. In the case of the 4d, codimension-1 dCFT which is dual to the D3-D5 probe brane system, the tree level one-point function of (length $\Delta = 2\ell$) chiral primary operators \eqref{CPOdefinition} is \cite{NagasakiYamaguchi12}:
\begin{IEEEeqnarray}{c}
\langle O(x) \rangle_{\text{tree}} = C_\ell \frac{(2\pi^2)^\ell}{\sqrt{2\ell}\lambda^\ell} \frac{k(k^2-1)^\ell}{x_3^{2 \ell}}, \ C_{\ell} \equiv \frac{1}{2^{\ell}}\sqrt{\frac{\ell+1}{2\ell+1}}. \qquad \label{OnePointFunctionCPOweak}
\end{IEEEeqnarray}
\indent In the case of the ferromagnetic vacuum state operator \eqref{VacuumState}, we can go one step further. The tree level one-point function is given by \cite{deLeeuwKristjansenZarembo15, Buhl-MortensenLeeuwIpsenKristjansenWilhelm16a}:
\begin{IEEEeqnarray}{c}
\langle \mathcal{O}(x) \rangle_{\text{tree}} = \frac{1}{2\ell+1}\sqrt{\frac{2}{\ell}}\left(\frac{8\pi^2}{\lambda}\right)^{\ell}\cdot \frac{B_{2\ell+1}\left(\frac{1+k}{2}\right)}{x_3^{2\ell}}, \qquad \label{OnePointFunctionVacuumTree}
\end{IEEEeqnarray}
where $B_n(x)$ are the Bernoulli polynomials and $L = 2\ell$ is the length of the operator ($\ell = 0,1,\ldots$). Both one-point functions \eqref{OnePointFunctionCPOweak}, \eqref{OnePointFunctionVacuumTree} vanish when the lengths $\Delta$ and $L$ are odd. Expanding the one-point function \eqref{OnePointFunctionVacuumTree} around $k = \infty$, we are led to the following perturbative expansion:
\begin{IEEEeqnarray}{ll}
\langle \mathcal{O}(x) \rangle_{\text{tree}} &= \frac{1}{2\ell+1}\cdot\frac{(2\pi^2)^\ell}{\sqrt{2\ell} \, \lambda^\ell} \ \frac{(k+1)^{2\ell+1}}{x_3^{2 \ell}} \times \nonumber \\ 
&\times\left[1 - \frac{2\ell+1}{k} + \frac{(2\ell+1)(2\ell+3)}{3k^2} - \ldots\right], \qquad
\end{IEEEeqnarray}
which agrees with the large-$k$ behavior of the CPO tree level one-point function \eqref{OnePointFunctionCPOweak}. For the vacuum state operator \eqref{VacuumState}, the one-loop correction to the one-point function has also been computed \cite{Buhl-MortensenLeeuwIpsenKristjansenWilhelm16a, Buhl-MortensenLeeuwIpsenKristjansenWilhelm16c}:
\begin{IEEEeqnarray}{c}
\langle \mathcal{O}(x) \rangle_{\text{1-loop}} = \frac{\ell}{2\ell-1} \sqrt{\frac{2}{\ell}}\left(\frac{8\pi^2}{\lambda}\right)^{\ell-1}\cdot\frac{B_{2\ell-1}\left(\frac{1+k}{2}\right)}{x_3^{2\ell}}. \qquad \label{OnePointFunctionVacuumLoop}
\end{IEEEeqnarray}
Similar results have been obtained for the (codimension-1) D3-D7 dCFT \cite{KristjansenSemenoffYoung12b, deLeeuwKristjansenLinardopoulos16, GimenezGrauKristjansenVolkWilhelm18, GimenezGrauKristjansenVolkWilhelm19} and the (codimension-3) D1-D3 dCFT \cite{KristjansenZarembo23, KristjansenZarembo24b} in 4d. For the (codimension-1) beta-deformed D3-D5 dCFT in 4d, D2-D4 dCFT in 3d and (codimension-2) D3-D3 dCFT in 4d, only scalar tree-level one-point functions are currently available \cite{Widen18, KristjansenVuZarembo21, ChalabiKristjansenSu25}. See \cite{Widen17, deLeeuwIpsenKristjansenVardinghusWilhelm17} for scalar tree-level D3-D5 two-point functions, and \cite{deLeeuwKristjansenLinardopoulosVolk23, Linardopoulos25b} for spinorial tree-level two-point functions in the D3-D5 and D3-D7 dCFTs.
\subsection{Strong coupling \label{SubSection:TopDownStrong}}
\noindent AdS/{\color{Red}d}CFT correlation functions can also be computed on the dual string theory side. The near-horizon geometry of a large number $N$ of coincident D3-branes is AdS$_5 \times\tn{S}^5$ whose metric (in Euclidean signature) is,
\begin{IEEEeqnarray}{c}
ds^2 = \frac{R^2}{z^2} \left(dx_0^2 + dx_1^2 + dx_2^2 + dx_3^2 + dz^2\right) + R^2 d\Omega_5^2. \qquad \label{AdS5xS5metric}
\end{IEEEeqnarray}
The probe D5-brane extends in the radial direction of AdS$_5$ and wraps AdS$_4\times\text{S}^2\subset\text{AdS}_5\times\text{S}^5$, with $k$ units of magnetic flux through S$^2$. Solving the probe D5-brane equations of motion, we find the following solution \cite{KarchRandall01b}:
\begin{IEEEeqnarray}{c}
x_3 = \kappa\cdot z, \qquad \kappa = \frac{\pi k}{\sqrt{\lambda}}. \label{D5braneEmbedding}
\end{IEEEeqnarray}
\indent One-point functions can be computed with the Gubser-Klebanov-Polyakov-Witten (GKPW) prescription \cite{GubserKlebanovPolyakov98, Witten98a}. For the CPO \eqref{CPOdefinition}, the result takes the following form \cite{NagasakiYamaguchi12, Buhl-MortensenLeeuwIpsenKristjansenWilhelm17a} (omitting higher-order terms, see \cite{GeorgiouLinardopoulosZoakos23, Linardopoulos25a} for more):
\begin{IEEEeqnarray}{ll}
\langle O(x) \rangle &= \frac{C_{\ell}}{\Delta-1} \sqrt{\frac{\lambda}{2^{\Delta}\pi^2\Delta}}\cdot\frac{1}{x_3^{\Delta}}\cdot\left(\Delta\sqrt{1 + \kappa^2} - \kappa\right) \times \qquad \nonumber\\[6pt]
 &\times\left(\sqrt{1 + \kappa^2}+\kappa \right)^\Delta = C_\ell\sqrt{\frac{2^{\Delta}\lambda}{\pi^2\Delta}} \ \frac{\kappa^{\Delta+1}}{x_3^{\Delta}}\times \nonumber\\[6pt]
&\times\left(1 + \frac{\Delta(\Delta+1)}{4(\Delta-1)\kappa^2} + \frac{\Delta(\Delta+1)}{32\kappa^4} + \ldots\right) = \nonumber\\[9pt]
& = C_\ell\,\frac{(2\pi^2)^\ell}{\sqrt{2\ell} \, \lambda^\ell} \ \frac{k^{2\ell+1}}{x_3^{2 \ell}} + \ldots, \qquad \label{OnePointFunctionCPOstrong1}
\end{IEEEeqnarray}
where $\Delta = 2\ell$ as before and the ellipsis denotes higher-order corrections in $1/\kappa$, for $\kappa \gg 1$. The leading (in $1/\kappa$) term of the strong coupling result \eqref{OnePointFunctionCPOstrong1} exactly reproduces the weak coupling, tree level result \eqref{OnePointFunctionCPOweak} in the double scaling limit $k \gg \sqrt{\lambda} \gg 1$, $\kappa \gg 1$. This limit is very reminiscent of the BMN limit \cite{BMN02} whereby the perturbative expansion of anomalous dimensions of unprotected operators of $\N = 4$ SYM organizes itself (up to three loops) in powers of $\lambda/J^2$, $J$ being the angular momentum of the operator. Interestingly, the strong coupling value \eqref{OnePointFunctionCPOstrong1} of the one-point function of the CPO \eqref{CPOdefinition} can be turned into a nontrivial test for the one-loop correction \eqref{OnePointFunctionVacuumLoop} to the one-point function of the vacuum state operator \eqref{VacuumState} \cite{Buhl-MortensenLeeuwIpsenKristjansenWilhelm16a, Buhl-MortensenLeeuwIpsenKristjansenWilhelm16c} since, for $\Delta = L = 2\ell$,
\begin{IEEEeqnarray}{c}
\left.\frac{\langle O(x)\rangle_{\text{1-loop}}}{\langle O(x) \rangle_{\text{tree}}}\right|_{\text{string}} = \left.\frac{\langle \mathcal{O}(x) \rangle_{\text{1-loop}}}{\langle \mathcal{O}(x)\rangle_{\text{tree}}}\right|_{\text{gauge}} = \nonumber \\[6pt]
= \frac{\lambda}{4\pi^2k^2}\frac{L(L+1)}{L-1}, \label{OnePointFunctionTest}
\end{IEEEeqnarray}
despite the fact that the operators \eqref{CPOdefinition} and \eqref{VacuumState} are not the same. This successful test \eqref{OnePointFunctionTest} further leads to yet another prediction for the asymptotic form of the perturbative expansion at higher loop orders \cite{Buhl-MortensenLeeuwIpsenKristjansenWilhelm17a}. In the heavy operator limit, which is obtained by sending $\Delta,\ell \to \infty$ while keeping $\kappa^2 \gg \Delta$, the strong coupling (tree level) one-point function \eqref{OnePointFunctionCPOstrong1} of the CPO \eqref{CPOdefinition} takes the form,
\begin{IEEEeqnarray}{c}
\langle O(x) \rangle = C_\ell\sqrt{\frac{2^{\Delta}\lambda}{\pi^2\Delta}} \ \frac{\kappa^{\Delta+1}}{x_3^{\Delta}} \underbrace{\left(1 + \frac{\Delta}{4\kappa^2} + \frac{\Delta^2}{32\kappa^4} + \ldots\right)}_{\exp\left(\Delta/4\kappa^2\right)}. \qquad \label{OnePointFunctionCPOstrong2}
\end{IEEEeqnarray}
\indent Similar results have been found for the D3-D7, D2-D4 and D1-D3 probe-brane systems \cite{KristjansenSemenoffYoung12b, GimenezGrauKristjansenVolkWilhelm18, GimenezGrauKristjansenVolkWilhelm19, KristjansenVuZarembo21, KristjansenZarembo23, KristjansenZarembo24b}. Scalar two-point functions of the D3-D5 probe-brane system have been computed in \cite{GeorgiouLinardopoulosZoakos23}; see \cite{GeorgiouLinardopoulosZoakos25} for the stringy computation of a spinorial one-point function. \\
\indent The one-point function of the ferromagnetic vacuum state operator \eqref{VacuumState} has been computed with semiclassical strings \cite{Buhl-MortensenLeeuwKristjansenZarembo15}, in the double scaling ($k \gg \sqrt{\lambda} \gg 1$, $\kappa \gg 1$), heavy operator (BMN) limit ($L\sim\sqrt{\lambda}\sim\sqrt{N}$). The result reads:
\begin{IEEEeqnarray}{ll}
\langle \mathcal{O}(x) \rangle = \frac{\left(\sqrt{1+\kappa^2} + \kappa\right)^L}{\left(\sqrt{2}x_3\right)^L} = &\frac{2^{L/2}\kappa^L}{x_3^L}\bigg[1 + \frac{L}{4\kappa^2} + \nonumber \\
&+ \frac{L(L-3)}{32\kappa^4} + \ldots\bigg], \qquad \label{OnePointFunctionVacuumStrong}
\end{IEEEeqnarray}
and it turns out to be the same with the one derived with the geodesic approximation method as we will see below. 

\newpage\section{Bottom-up approach \label{Section:BottomUpApproach}}
\noindent In the bottom-up approach, the codimension-1 probe brane $\Sigma$ is viewed as an \quotes{interface} or \quotes{domain wall}, i.e.\ a dynamical boundary that hosts gravitational degrees of freedom. The total action of the system is given by the sum of the bulk Einstein-Hilbert (EH) action, the boundary improvement term of Gibbons-Hawking-York (GHY) \cite{York72b, GibbonsHawking76} and the action of the domain wall:
\begin{IEEEeqnarray}{c}
S = S_{\text{EH}} + S_{\text{GHY}} + S_{\text{brane}}. \label{ActionDW}
\end{IEEEeqnarray}
By varying the action \eqref{ActionDW} we obtain Einstein's field equations in the bulk, along with the Israel matching conditions on the boundary \cite{ChamblinReall99}:
\begin{IEEEeqnarray}{c}
\gamma^{I}_{\mu\nu} = \gamma^{II}_{\mu\nu}, \qquad \pi_{\mu\nu}^{I} - \pi_{\mu\nu}^{II} = t_{\mu\nu}, \label{IsraelMatchingConditions}
\end{IEEEeqnarray}
where $\gamma_{\mu\nu}^a$ is the first fundamental form, $t_{\mu\nu}$ is the energy-momentum tensor of the brane, and $\pi_{\mu\nu}^a$ is the canonical momentum,
\begin{IEEEeqnarray}{c}
t_{\mu\nu} \equiv \frac{2}{\sqrt{-\gamma}}\cdot\frac{\delta S_{\text{brane}}}{\delta \gamma^{\mu\nu}}, \qquad \pi^a_{\mu\nu} \equiv K^a_{\mu\nu} - K^a\gamma^a_{\mu\nu}, \qquad \label{EMT_Momentum}
\end{IEEEeqnarray}
which is defined in terms of the extrinsic curvature $K^a_{\mu\nu}$ and its trace $K^a$ on each side of the domain wall ($a = \pm$). See appendix \ref{AppendixB} for the corresponding definitions. The embedding of the probe brane can be determined by solving the junction equation which follows from \eqref{IsraelMatchingConditions}. \\
\indent Let us consider a 4-dimensional interface brane inside AdS$_5$. The brane separates AdS$_5$ into two subregions AdS$_I$ and AdS$_{II}$, one for each copy of $\N = 4$ SYM on the dual field theory side. For consistency with the top-down approach, the interface brane intersects the AdS boundary (located at $z = 0$, according to the Poincaré metric \eqref{AdS5xS5metric}) at $x_3 = 0$. This way, the AdS$_I$ subregion corresponds to CFT$_I$ to the right of the defect ($x_3 > 0$) and the AdS$_{II}$ subregion corresponds to CFT$_{II}$ to left of the defect ($x_3 < 0$). See figure \ref{Figure1} below. \\
\begin{figure}[H]\begin{center}\begin{tikzpicture}[scale=1]
\draw[->] (-3.75,0) to (3.75,0);
\draw[->] (0,0) to (0,3);
\node at (-0.2,3) {$z$};
\node[] at (3.75,-0.3) {$x_3$};
\node[] at (2.7,-0.3) {CFT$_{\text{I}}$};
\node[] at (-1.7,-0.3) {CFT$_{\text{II}}$};
\node[] at (3.4,.7) {AdS$_{\text{I}}$};
\node[] at (-1.5,1.8) {AdS$_{\text{II}}$};
\node[] at (3,2.3) {\begin{tabular}{c} interface \vspace{-.2cm} \\ brane \end{tabular}};
\draw[-,Red,thick] (0,0) to (3.75,2);
\draw[-,thick,blue] (3,0) arc (0:28:2.97);
\draw[-,thick,violet] (-1.5,0) arc (0:-142:-2.31);
\end{tikzpicture}\caption{Interface brane in the bottom-up model.}\label{Figure1}\end{center}\end{figure}
\indent Because the two copies of $\N = 4$ SYM have different gauge groups ($SU_{I}(N)$ on the right, $SU_{II}(N-k)$ on the left) and 't Hooft couplings ($\lambda_I \equiv \gym^2 N$, $\lambda_{II} \equiv \gym^2 (N-k)$), their dual AdS subregions will have different radii:
\begin{IEEEeqnarray}{c}
R_I^2 = \alpha'\sqrt{\gym^2 N} \qquad \& \qquad R_{II}^2 = \alpha'\sqrt{\gym^2 (N-k)}, \qquad
\end{IEEEeqnarray}
where $\gym^2 \equiv 4\pi g_s$. To determine the embedding of the interface brane inside AdS$_5$, we assume that the energy-momentum tensor is of the following form:
\begin{IEEEeqnarray}{c}
t_{\mu\nu} = \sigma\cdot\gamma_{\mu\nu}, \label{EMT_Ansatz}
\end{IEEEeqnarray}
where $\sigma$ stands for the tension of the interface brane. It can be shown that the tension $\sigma$ is related to the entropy of the corresponding boundary state \cite{AffleckLudwig91d, AzeyanagiKarchTakayanagiThompson07}. As such it can be considered as a measure of the boundary degrees of freedom. When $\sigma$ is a constant, the form \eqref{EMT_Ansatz} of the energy-momentum tensor is consistent with a constant domain wall Lagrangian \cite{Takayanagi11b, FujitaTakayanagiTonni11}. By solving the Israel matching conditions \eqref{IsraelMatchingConditions}, the following solution can be found (see e.g.\ \cite{CzechNguyenSwaminathan16} and references therein):
\begin{IEEEeqnarray}{c}
x_3^a = \kappa^a\cdot z^a, \qquad R_I\sqrt{1+\kappa_I^2} = R_{II}\sqrt{1+\kappa_{II}^2}, \qquad \label{BraneEmbeddingBackreacting}
\end{IEEEeqnarray}
where the slope of the interface brane jumps across the defect.\\
\indent To keep the same brane slope as we move across the defect, we should give up the first Israel matching condition in \eqref{IsraelMatchingConditions}. Specifically, we allow the first fundamental form $\gamma_{\mu\nu}$ to rescale by a constant overall factor, $\gamma_{\mu\nu}^I = (R^2_I/R^2_{II})\cdot\gamma_{\mu\nu}^{II}$. Considering the top-down (probe-brane) ansatz \cite{KarchRandall01b} for the embedding of the interface brane,
\begin{IEEEeqnarray}{c}
z = z(x_3), \label{BraneEmbedding1}
\end{IEEEeqnarray}
we arrive at the following matching condition:
\begin{IEEEeqnarray}{c}
\frac{3}{\sqrt{1 + \acute{z}^2}}\cdot\left(\frac{1}{r_+} - \frac{1}{r_-}\right) = \sigma, \quad r_{\pm} \equiv R_{\pm} + \frac{R_{\mp}^7}{R_{\pm}^6}, \qquad \label{JunctionEquation1}
\end{IEEEeqnarray}
where $R_+ = R_I$ and $R_- = R_{II}$, and primes denote differentiation wrt $x_3$ ($\acute{z} \equiv \partial_3 z$). The details of the derivation can be found in appendix \ref{AppendixB}. Solving \eqref{JunctionEquation1}, we find the following configuration of the interface brane:
\begin{IEEEeqnarray}{c}
x_3 = \kappa\cdot z, \qquad \kappa = \frac{\sigma \, r_+ r_-}{\sqrt{9\left(r_+ - r_-\right)^2 - \sigma^2r_+^2 r_-^2}}. \qquad \label{BraneEmbeddingBottomUp}
\end{IEEEeqnarray}
For consistency with the probe-brane result \eqref{D5braneEmbedding}, the tension of the interface brane should be set equal to,
\begin{IEEEeqnarray}{c}
\sigma = \frac{3\kappa}{\sqrt{1 + \kappa^2}}\cdot\frac{\left|r_+ - r_-\right|}{r_+ r_-}. \label{BraneTension}
\end{IEEEeqnarray}
In computing correlation functions below, we mainly focus on one side of the domain wall (namely side $I$), so that our results are for the most part independent of the explicit value of $\kappa$ (as specified from the probe-brane solution \eqref{D5braneEmbedding}, the backreacting solution \eqref{BraneEmbeddingBackreacting} or the solution \eqref{BraneEmbeddingBottomUp}). The backreaction of the brane to the geometry of spacetime is ignored so that all geodesics will be probe-brane geodesics. No geodesics cross the domain wall. A few computations on side $II$ of the domain wall pertain to solution \eqref{BraneEmbeddingBottomUp} for reference purposes only.
\subsection{One-point functions \label{SubSection:OnePointFunctions}}
\noindent One-point functions (expectation values) of local operators vanish in CFTs, whereas two and higher-point functions are generally nonzero. As we have seen in the introduction, the correlation function of two local operators on the boundary of AdS can be computed (up to an overall normalization) from the length of the geodesic curve which connects them in the bulk \cite{SusskindWitten98, Solodukhin98b, BalasubramanianRoss99, LoukoMarolfRoss00, KrausOoguriShenker03}:
\begin{IEEEeqnarray}{lll}
\langle \hat{O}(x)\hat{O}(y) \rangle \simeq e^{-\Delta\cdot L(x; y)/R},
\end{IEEEeqnarray}
where $\Delta$ is the scaling dimension of the local operator $\hat{O}(x)$, and $L(x; y)$ is the geodesic distance of the boundary points $x$ and $y$. The unrenormalized operators $\hat{O}(x)$ can be renormalized by absorbing their UV divergences as ($\epsilon$ is the short distance cutoff),
\begin{IEEEeqnarray}{c}
O(x) = \frac{\hat{O}(x)}{\epsilon^{\Delta}}.
\end{IEEEeqnarray}
From now on, we only consider renormalized operators by ignoring their UV divergences.\\
\indent In boundary and defect CFTs, one-point functions of local operators no longer vanish in the general case. For codimension-1 boundaries and defects, only scalar operators have non-vanishing one-point functions. In this case, one-point functions can again be determined from the geodesic distance between the operator on the boundary of AdS and the codimension-1 boundary/defect. Focusing on the AdS$_5$/{\color{Red}d}CFT$_4$ correspondence with a 3-dimensional (codimension-1) planar defect situated at the point $x_3 = 0$ of 4-dimensional ambient space, the one-point function of a (renormalized) local scalar operator is given, up to an overall normalization $\NN$, by
\begin{IEEEeqnarray}{c}
\langle O(x) \rangle \equiv \langle O(\textbf{x},x_3)|B\rangle = \NN\cdot\frac{e^{-\Delta \cdot L(\textbf{x}, x_3; \textbf{x}_r, x_r)/R}}{\epsilon^{\Delta}}, \qquad \label{OnePointFunctionGeodesic}
\end{IEEEeqnarray}
where $x \equiv (\textbf{x},x_3)$, $\textbf{x} \equiv (x_0,x_1,x_2)$, while $(\textbf{x}_r, x_r, z_r)$ is a reflection point on the interface brane which minimizes the distance to the location $(\textbf{x}, x_3)$ of the operator $O(x)$ on the boundary of AdS. In the case of CFT$_I$ ($x_3 > 0$), the geodesic distance from the AdS boundary to the brane is,
\begin{IEEEeqnarray}{c}
L(\textbf{x},x_3;\textbf{x}_r,x_r) = R_I \int_{x_r}^{x_3} d\tilde{x}_3 \cdot \frac{\sqrt{1 + \acute{\textbf{x}}^2 + \acute{z}^2}}{z}, \qquad \label{GeodesicDistance1}
\end{IEEEeqnarray}
where once more, primes denote derivatives wrt $x_3$ (and the corresponding integration variable $\tilde{x}_3$), i.e.\ $\acute{\textbf{x}} \equiv \partial_3\textbf{x}$ and $\acute{z} \equiv \partial_3z$. Using the conserved quantities of the integrand \eqref{GeodesicDistance1}, the geodesic distance \eqref{GeodesicDistance1} can be computed analytically to be \cite{KimPalPark23, Park24}:
\begin{IEEEeqnarray}{c}
L(x;\textbf{x}_r,x_r) = R_I \log\left[\frac{|\textbf{x}-\textbf{x}_r|^2 + |x_3-x_r|^2 + z_r^2}{z_r\,\epsilon}\right], \qquad \label{GeodesicDistance2}
\end{IEEEeqnarray}
where $\epsilon$ is the UV cutoff and $x_r = \kappa\cdot z_r$ (where $\kappa$ is given by \eqref{BraneEmbeddingBottomUp}), because the reflection point lies on the interface brane. By varying the geodesic distance in \eqref{GeodesicDistance2} wrt $\textbf{x}_r$ and $z_r$ (having also used $x_r = \kappa\cdot z_r$) we can determine the location of the reflection point:
\begin{IEEEeqnarray}{c}
\textbf{x}_r = \textbf{x} \qquad \& \qquad z_r = \frac{x_3}{\sqrt{1 + \kappa^2}} > 0. \label{ReflectionPoint1}
\end{IEEEeqnarray}
Plugging this solution, first into the formula for the geodesic distance \eqref{GeodesicDistance2},
\begin{IEEEeqnarray}{c}
L(\textbf{x},x_3;\textbf{x}_r,x_r) = R_I \log\bigg(\frac{2x_3(\sqrt{1+\kappa^2} - \kappa)}{\epsilon}\bigg), \qquad \label{GeodesicDistance3}
\end{IEEEeqnarray}
and then into the expression \eqref{OnePointFunctionGeodesic} for the holographic one-point function, we find:
\begin{IEEEeqnarray}{c}
\langle O(x) \rangle_{I} \equiv \langle O(\textbf{x},x_3)|I\rangle = \frac{{\CC}_I}{x_3^{\Delta}}. \label{OnePointFunctionI1}
\end{IEEEeqnarray}
The one-point function structure constant \cite{KastikainenShashi21} and its $\kappa \gg 1$ perturbative expansion are identical to the top-down result \eqref{OnePointFunctionVacuumStrong} for the one-point function of the ferromagnetic vacuum state operator:\footnote{We thank C.\ Kristjansen for bringing the paper \cite{KastikainenShashi21} to our attention, as well as for pointing out to us the computation of the one-point function \eqref{OnePointFunctionVacuumStrong} which was carried out in \cite{Buhl-MortensenLeeuwKristjansenZarembo15}.}
\begin{IEEEeqnarray}{ll}
\CC_I = &\frac{\NN_I}{2^{\Delta}\left(\sqrt{1 + \kappa^2} - \kappa\right)^{\Delta}} = \NN_I\cdot\kappa^{\Delta}\bigg(1 + \frac{\Delta}{4\kappa^2} + \nonumber\\[6pt]
&+ \frac{\Delta(\Delta-3)}{32\kappa^4} + \frac{\Delta(\Delta-4)(\Delta-5)}{384\kappa^6} + \cdots \bigg), \qquad \label{OnePointFunctionI2}
\end{IEEEeqnarray}
where the overall normalization factor $\NN_I$ includes the contribution of the compact internal space (originating in the scalar field sector that is dual to the 5-sphere) which does not partake in the geodesic approximation formalism \eqref{OnePointFunctionGeodesic} (cf.\ \cite{Park20a, Park21b, KimKimLeeParkSeo23, ParkKimCho24, Park24}). \\
\indent In the heavy operator limit $\Delta\rightarrow\infty$, the perturbative expansion of the bottom-up one-point function structure constant \eqref{OnePointFunctionI2} takes the following form, for $\kappa \gg \Delta \gg 1$:
\begin{IEEEeqnarray}{c}
\CC_I = \NN_I\cdot\kappa^{\Delta}\underbrace{\left(1 + \frac{\Delta}{4\kappa^2} + \frac{\Delta^2}{32\kappa^4} + \frac{\Delta^3}{384\kappa^6} + \cdots \right)}_{\exp\left(\Delta/4\kappa^2\right)}, \qquad \label{OnePointFunctionCPOstrong3}
\end{IEEEeqnarray}
which matches exactly the corresponding top-down result \eqref{OnePointFunctionCPOstrong2} for the one-point function of CPOs, as long as we identify,
\begin{IEEEeqnarray}{c}
\NN_I = C_\ell\,\kappa\,\sqrt{\frac{2^{\Delta}\lambda}{\pi^2\Delta}}.
\end{IEEEeqnarray}
\indent We can carry out a similar computation for the case of CFT$_{II}$. The operator is located at the point $(\textbf{y}, y_3)$, i.e.\ a transverse distance $y_3 < 0$ from the planar 3-dimensional defect at $y_3 = 0$. Repeating the above steps, we find the reflection point $(\textbf{y}_r, y_r, u_r)$,
\begin{IEEEeqnarray}{c}
\textbf{y}_r = \textbf{y} \qquad \& \qquad u_r = -\frac{y_3}{\sqrt{1 + \kappa^2}} > 0.
\end{IEEEeqnarray}
The one-point function is again given by,
\begin{IEEEeqnarray}{c}
\langle O(y) \rangle_{II} \equiv \langle O(\textbf{y},y_3)|II\rangle = \frac{{\CC}_{II}}{y_3^{\Delta}}, \label{OnePointFunctionII}
\end{IEEEeqnarray}
where $y \equiv (\textbf{y}, y_3)$, $\textbf{y} \equiv (y_0,y_1,y_2)$ and the structure constant, along with its $\kappa \gg 1$ expansion, read:
\begin{IEEEeqnarray}{ll}
\CC_{II} = &\frac{\NN_{II}}{2^{\Delta}\left(\sqrt{1 + \kappa^2} + \kappa\right)^{\Delta}} = \frac{\NN_{II}}{2^{2\Delta}\kappa^{\Delta}}\bigg(1 - \frac{\Delta}{4\kappa^2} + \nonumber \\[6pt]
& + \frac{(\Delta+3)\Delta}{32\kappa^4} - \frac{(\Delta+5)(\Delta+4)\Delta}{384\kappa^6} + \cdots \bigg). \qquad\quad
\end{IEEEeqnarray}
Note that the one-point function structure constants $\CC_I$ and $\CC_{II}$ satisfy,
\begin{IEEEeqnarray}{c}
\CC_{I}\CC_{II} = \frac{\NN_I\NN_{II}}{2^{2\Delta}}. \label{OnePointFunctionProduct1}
\end{IEEEeqnarray}
\subsection{Two-point functions \label{SubSection:TwoPointFunctions}}
\noindent We now turn our attention to two-point functions. As a warm-up, we compute the correlation function of two local scalar operators, $O_1(x)$ and $O_2(y)$ in the absence of a defect. Based on the recipe \eqref{TwoPointFunctionCFT1}, the (renormalized) two-point function can be computed from the formula,
\begin{IEEEeqnarray}{c}
\langle O_1(x)O_2(y)\rangle = \NN\cdot\frac{e^{-L(x,y)}}{\epsilon^{\Delta_1+\Delta_2}}, \label{TwoPointFunctionGeodesic1}
\end{IEEEeqnarray}
where $\NN$ is a normalization constant and $\Delta_{1,2}$ are the scaling dimensions of the operators. For equal scaling dimensions, $\Delta = \Delta_1 = \Delta_2$, the geodesic distance reads,
\begin{IEEEeqnarray}{c}
L(x,y) = \Delta\log\left(\frac{|x-y|^2}{\epsilon^2}\right),
\end{IEEEeqnarray}
which, upon inserting into the two-point function formula \eqref{TwoPointFunctionGeodesic1}, leads to the expected result:
\begin{IEEEeqnarray}{c}
\langle O(x,y)\rangle = \frac{1}{(x-y)^{2\Delta}}. \label{TwoPointFunctionCFT2}
\end{IEEEeqnarray}
\eqref{TwoPointFunctionCFT2} gives the correlation function of two ambient CFT operators with equal scaling dimensions $\Delta$, in the absence of a boundary/defect. \\
\indent The generic form of two-point functions in the case of codimension-1 boundary and defect CFTs is \eqref{TwoPointFunction}. For two scalar operators $O_1(\textbf{x},x_3)$ and $O_2(\textbf{y},y_3)$, which are placed at distances $x_3 > 0$ and $y_3 < 0$ from a planar codimension-1 defect at $x_3 = y_3 = 0$, their two-point correlation function can again be computed from \eqref{TwoPointFunctionGeodesic1}. The total geodesic distance between the operators on the boundary of AdS and the interface brane in the bulk is,
\begin{IEEEeqnarray}{c}
L(\textbf{x},x_3;\textbf{y},y_3) \equiv \sum_{a=1}^2 \frac{\Delta_a L_a}{R_a}, \label{GeodesicDistance4}
\end{IEEEeqnarray}
where $L_a$ ($a = 1,2$) is the geodesic distance of each operator from the interface brane, as defined in \eqref{GeodesicDistance1} above. The total geodesic distance \eqref{GeodesicDistance4} is given by,
\begin{IEEEeqnarray}{l}
L(\textbf{x},x_3;\textbf{y},y_3) = \Delta_1\log\left(\frac{|\textbf{x} - \textbf{x}_r|^2 + |x_3-x_r|^2 + z_r^2 }{z_r\,\epsilon}\right) + \nonumber \\[6pt]
\hspace{1.6cm} + \Delta_2\log\left(\frac{|\textbf{y}-\textbf{x}_r|^2 + |y_3-x_r|^2 + z_r^2}{z_r\,\epsilon}\right), \qquad \label{GeodesicDistance5}
\end{IEEEeqnarray}
where $(\textbf{x}_r,x_r,z_r)$ is the reflection point on the interface brane. The reflection point is common for the two operators and will be determined below by requiring that it minimizes the total geodesic distance \eqref{GeodesicDistance4}. Because the reflection point lies on the interface brane, $x_r = \kappa\cdot z_r$. \\
\indent For simplicity, we assume that the two operators have the same longitudinal coordinates $\textbf{x} = \textbf{y}$ and equal conformal dimensions, that is $\Delta_1 = \Delta_2 = \Delta$. By varying the geodesic length $L$ in \eqref{GeodesicDistance5} wrt both $\textbf{x}_r$ and $z_r$, we find the following expression for the common reflection point:
\begin{IEEEeqnarray}{l}
\textbf{x}_r = \textbf{x} = \textbf{y}, \nonumber \\[6pt]
z_r = \frac{\kappa(x_3+y_3)+\sqrt{\kappa^2(x_3-y_3)^2 - 4x_3y_3}}{2\left(1+\kappa^2\right)}.
\end{IEEEeqnarray}
Using this result, the two-point function \eqref{TwoPointFunctionGeodesic1} once more takes the form of a conformal two-point function, 
\begin{IEEEeqnarray}{c}
\langle O(\textbf{x},x_3)O(\textbf{x},y_3)\rangle = \frac{1}{|x_3 - y_3|^{2\Delta}}, \label{TwoPointFunctionGeodesic2}
\end{IEEEeqnarray}
where $x_3 - y_3 = x_3 + |y_3|$ is the distance of the two operators. The form of the correlator \eqref{TwoPointFunctionGeodesic2} implies that the defect obstructs the transmission of massive modes so that the fields in the scalar two-point function $\langle OO\rangle$ are Wick-contracted with the (massless) CFT propagators. \\
\indent Taking $x_3 = -y_3$, the scalar two-point function \eqref{TwoPointFunctionGeodesic2} can be written as the product of the one-point functions \eqref{OnePointFunctionI1}, \eqref{OnePointFunctionII} that we found above,
\begin{IEEEeqnarray}{ll}
\langle O(\textbf{x},x_3)O(\textbf{x},-x_3)\rangle &= \langle O(\textbf{x},x_3)|I\rangle \cdot \langle II| O(\textbf{x},-x_3)\rangle = \nonumber \\[6pt]
&= \frac{{\CC}_I}{|x_3|^{\Delta}}\cdot\frac{{\CC}_{II}}{|- x_3|^{\Delta}} = \frac{1}{|2x_3|^{2\Delta}}, \qquad \label{OnePointFunctionProduct2}
\end{IEEEeqnarray}
which gives the two-point function of a scalar local operator and its mirror image across the planar 3d defect at $x_3 = y_3 = 0$. Note that we also need to take $\NN_I\NN_{II} = 1$ for the normalization constants in \eqref{OnePointFunctionProduct1}. The two-point function \eqref{OnePointFunctionProduct2} is in fact a special case of the disconnected two-point function of the operators $O_1(\textbf{x},x_3)$ and $O_2(\textbf{y},y_3)$, evaluated at a point ($\textbf{x} = \textbf{y}$, $x_3 = -y_3$) where it coincides with the connected two-point function. \\
\indent That said, let us also write out the disconnected two-point function of two scalar operators $O_1(\textbf{x},x_3)$ and $O_2(\textbf{y},y_3)$, which are placed for instance to the right of the defect at $x = 0$ (side of CFT$_I$, with $x_3,y_3>0$):
\begin{IEEEeqnarray}{ll}
\langle O_1(\textbf{x},x_3)O_2(\textbf{y},y_3)\rangle &= \langle O_1(\textbf{x},x_3)|I\rangle \cdot \langle I| O_2(\textbf{y},y_3)\rangle = \nonumber \\[6pt]
& = \frac{\CC_1\CC_2}{|x_3|^{\Delta_1}|y_3|^{\Delta_2}}. \qquad
\end{IEEEeqnarray}
\subsubsection{Reflected two-point functions \label{SubSubSection:TwoPointReflected}}
\noindent We now compute reflected two-point functions. By reflected we mean that the fusion of two ambient operators occurs at the same point on the planar codimension-1 boundary. From the point of view of the bulk theory, two geodesics leave the location of the operators on the boundary of AdS and meet at the same reflection point on the interface brane. Because the general case turns out to be very complicated, we resort to two alternative ways to perform the computation. In the first (\quotes{collinear}) case, the two operators lie on a single straight line from the defect, while in the second (\quotes{equidistant}) case, the two operators lie at the same normal distance from the defect. See figure \ref{Figure2} below. Note that figure \ref{Figure2} applies to both the defect CFT and the bulk of AdS.
\begin{figure}[h]\begin{center}\begin{tikzpicture}\begin{feynman}
\fill [lightgray] (-.785,-.02) rectangle (+0.785,-.22);
\vertex (a) [dot] at (0,0) {};
\vertex (b) [empty dot] at (0,.7) {};
\vertex (c) [empty dot] at (0,1.4) {};
\vertex (f1) at (-.9,0) {};
\vertex (f2) at (+.9,0) {};
\vertex (O1) at (.4,.7) {$O_2$};
\vertex (O2) at (.4,1.4) {$O_1$};
\diagram* [line width=0.3mm, Red] {(c) -- (a) -- (b)};
\diagram* [line width=0.3mm, black] {(f1) -- (f2)};
\fill [lightgray] (3.3-.785,-.02) rectangle (3.3+0.785,-.22);
\vertex (a) [dot] at (3.3,0) {};
\vertex (b) [empty dot] at (3.3-.5,.7) {};
\vertex (c) [empty dot] at (3.3+.5,.7) {};
\vertex (f1) at (3.3-.9,0) {};
\vertex (f2) at (3.3+.9,0) {};
\vertex (O1) at (3.3-.7,1) {$O_1$};
\vertex (O2) at (3.3+.7,1) {$O_2$};
\diagram* [line width=0.3mm, Red] {(c) -- (a) -- (b)};
\diagram* [line width=0.3mm, black] {(f1) -- (f2)};
\end{feynman}\end{tikzpicture}\caption{Two ways to compute reflected two-point functions: collinear (left) and equidistant (right).}\label{Figure2}\end{center}\end{figure}
\paragraph{Collinear case}\hspace{-0.2cm}($\textbf{x} = \textbf{y}$)\hspace{1em} We first consider the case where the two operators lie to the right of the defect (i.e.\ on the CFT$_I$ side), at distances $x_3>y_3>0$ and are collinear, i.e.\ $\textbf{x} = \textbf{y}$. See the left drawing in figure \ref{Figure2}. The geodesic distance between the two operators and the interface brane is given as before by,
\begin{IEEEeqnarray}{ll}
L(\textbf{x},x_3,y_3) = &\Delta\bigg[\log\left(\frac{|\textbf{x} - \textbf{x}_r|^2 + |x_3-x_r|^2 + z_r^2 }{z_r\,\epsilon}\right) + \nonumber \\[6pt]
& + \log\left(\frac{|\textbf{x}-\textbf{x}_r|^2 + |y_3-x_r|^2 + z_r^2}{z_r\,\epsilon}\right)\bigg], \qquad
\end{IEEEeqnarray}
where $(\textbf{x}_r, x_r = \kappa\cdot z_r, z_r)$ is the reflection point in the bulk of AdS$_5$. By varying the geodesic length wrt $\textbf{x}_r$ and $z_r$, we find that the reflection point is given by,
\begin{IEEEeqnarray}{c}
\textbf{x}_r = \textbf{y} = \textbf{x}, \qquad \& \qquad z_r = \sqrt{\frac{x_3 y_3}{1 + \kappa^2}}.
\end{IEEEeqnarray}
The corresponding reflected defect CFT two-point function becomes:
\small\begin{IEEEeqnarray}{c}
\langle O(\textbf{x},x_3) O(\textbf{x},y_3)\rangle = \frac{1}{\left(\sqrt{1 + \kappa^2}(x_3 + y_3) - 2\kappa\sqrt{x_3 y_3}\right)^{2\Delta}}. \qquad \label{TwoPointFunctionGeodesic3}
\end{IEEEeqnarray}\normalsize
Substituting the value of the invariant ratio $\xi$,
\begin{IEEEeqnarray}{c}
1 + \xi = \frac{(x_3 + y_3)^2}{4|x_3||y_3|},
\end{IEEEeqnarray}
which has been defined in \eqref{TwoPointFunction}, the collinear reflected two-point function can be written as follows \cite{KastikainenShashi21},
\begin{IEEEeqnarray}{c}
\langle O(\textbf{x},x_3) O(\textbf{x},y_3)\rangle = \frac{\left(\sqrt{(1 + \xi)(1 + \kappa^2)} - \kappa\right)^{-2\Delta}}{(4x_3y_3)^{\Delta}}. \qquad \label{TwoPointFunctionGeodesic4}
\end{IEEEeqnarray}
Expanding the two-point function in the limit of small $\xi$ we find,
\begin{IEEEeqnarray}{l}
\langle O(\textbf{x},x_3) O(\textbf{x},y_3)\rangle = \frac{\CC_I^2}{(x_3y_3)^{\Delta}}\cdot\Big[1 - \frac{\Delta}{2}\left(1 + c^2\right)\xi + \nonumber \\[6pt]
\hspace{.7cm} + \frac{\Delta}{16}\left(1 + c^2\right)\left(2\left(1 + c^2\right)\Delta + 3 + c^2\right)\xi^2 + \ldots\Big], \qquad \label{TwoPointFunctionGeodesic5}
\end{IEEEeqnarray}
where we have defined (for $\NN_I = 1$),
\begin{IEEEeqnarray}{l}
c \equiv \frac{1}{\sqrt{1 + \kappa^2} - \kappa} = \sqrt{1 + \kappa^2} + \kappa = 2\,\CC_I^{1/\Delta}. \label{DefinitionC}
\end{IEEEeqnarray}
In the limit where the two points coincide, $y_3 \rightarrow x_3$, $\xi \rightarrow 0$, the reflected two-point function \eqref{TwoPointFunctionGeodesic5} reduces to the square of the one-point function in \eqref{OnePointFunctionI1}--\eqref{OnePointFunctionI2}, in agreement with the boundary operator expansion in \eqref{TwoPointFunctionBOE} when there are no ambient-boundary correlations ($\BB_{\bullet j} = 0$).
\paragraph{Equidistant case ($x_3 = y_3$)} In the equidistant case (right drawing in figure \ref{Figure3}), $x_3 = y_3$ and $\textbf{x} \neq \textbf{y}$. The geodesic length which determines $\langle O(\textbf{x},x_3)O(\textbf{y},x_3)\rangle$ becomes,
\begin{IEEEeqnarray}{ll}
L(\textbf{x},x_3,\textbf{y}) = &\Delta\bigg[\log\left(\frac{|\textbf{x} - \textbf{x}_r|^2 + |x_3-x_r|^2 + z_r^2 }{z_r\,\epsilon}\right) + \nonumber \\[6pt]
&+ \log\left(\frac{|\textbf{y}-\textbf{x}_r|^2 + |x_3-x_r|^2 + z_r^2}{z_r\,\epsilon}\right)\bigg]. \qquad \label{GeodesicDistance6}
\end{IEEEeqnarray}
The reflection point $(\textbf{x}_r, x_r, z_r)$ once more satisfies $x_r = \kappa\cdot z_r$, so that by extremizing the geodesic distance \eqref{GeodesicDistance6} wrt $\textbf{x}_r$ and $z_r$ we are led to the solution,
\begin{IEEEeqnarray}{c}
\textbf{x}_r = \frac{\textbf{x} + \textbf{y}}{2} \qquad \& \qquad z_r = \sqrt{\frac{4x_3^2 + (\textbf{x}-\textbf{y})^2}{4(1 + \kappa^2)}}, \qquad
\end{IEEEeqnarray}
which we plug into \eqref{GeodesicDistance6} and the two-point function recipe \eqref{TwoPointFunctionGeodesic1} obtaining
\begin{IEEEeqnarray}{c}
\langle O(\textbf{x},x_3)O(\textbf{y},x_3)\rangle = \frac{\left(\sqrt{(1+\xi)(1+\kappa^2)} - \kappa\right)^{-2\Delta}}{(2x_3)^{2\Delta}}, \qquad \label{TwoPointFunctionGeodesic6}
\end{IEEEeqnarray}
which is nothing more than the collinear result \eqref{TwoPointFunctionGeodesic4}, for $y_3 = x_3$ and
\begin{IEEEeqnarray}{c}
\xi = \frac{(\textbf{x}-\textbf{y})^2}{4x_3^2}.
\end{IEEEeqnarray}
The small $\xi$ expansion of \eqref{TwoPointFunctionGeodesic6} is obviously very similar to \eqref{TwoPointFunctionGeodesic5}.
\subsubsection{Ambient-channel two-point functions \label{SubSubSection:TwoPointAmbient}}
\noindent In the ambient channel, the operators interact with the planar codimension-1 boundary via ambient light modes, that is primary operators of the ambient CFT. See figure \ref{Figure3} below. The two operators fuse at a 3-point vertex by emitting ambient light modes which mediate the interaction with the boundary. To compute defect two-point functions in the ambient channel we need, apart from a reflection point on the interface brane, an additional junction point in the bulk of AdS where three geodesic curves meet (described by the same figure \ref{Figure3}).
\begin{figure}[h]\begin{center}\begin{tikzpicture}\begin{feynman}
\fill [lightgray] (-.785,-1) rectangle (+0.785,-1.2);
\vertex (a) [dot] at (0,0) {};
\vertex (b) [empty dot] at (-.5,.7) {};
\vertex (c) [empty dot] at (+.5,.7) {};
\vertex (d) [dot] at (0,-.98) {};
\vertex (f1) at (-.9,-.98) {};
\vertex (f2) at (+.9,-.98) {};
\vertex (O1) at (-.7,1) {$O_1$};
\vertex (O2) at (+.7,1) {$O_2$};
\vertex (Oj) at (+.3,-.5) {\color{Blue}$O_j$};
\diagram* [line width=0.3mm, Red] {(c) -- (a) -- (b)};
\diagram* [line width=0.3mm, Blue] {(a) -- (d)};
\diagram* [line width=0.3mm, black] {(f1) -- (f2)};
\end{feynman}\end{tikzpicture}\caption{Ambient-channel two-point function.}\label{Figure3}\end{center}\end{figure}
One part of the geodesic curve connects the bulk junction point $(\textbf{x}_j, x_j, z_j)$ to the reflection point $(\textbf{x}_r, x_r, z_r)$ on the interface brane. This part is described by the following bulk-to-bulk propagator:
\begin{widetext}\begin{IEEEeqnarray}{c}
L_{jr} = 2R\log\left[\frac{\sqrt{(\textbf{x}_j - \textbf{x}_r)^2 + (x_j - x_r)^2 + (z_j - z_r)^2} + \sqrt{(\textbf{x}_j - \textbf{x}_r)^2 + (x_j - x_r)^2 + (z_j + z_r)^2}}{2\sqrt{z_j z_r}} \right]. \qquad \ \label{GeodesicDistance7a}
\end{IEEEeqnarray}\end{widetext}
The other part of the geodesic curve consists of two bulk-to-boundary propagators which emanate from the junction point $(\textbf{x}_j, x_j, z_j)$ and terminate at the locations ($(\textbf{x}, x_3, 0)$ and $(\textbf{y}, y_3, 0)$) of the local operators on the boundary of AdS$_5$:
\begin{IEEEeqnarray}{l}
L_1 = R\log\left(\frac{(\textbf{x} - \textbf{x}_j)^2 + (x_3 - x_j)^2 + z_j^2}{z_j\epsilon}\right), \qquad \label{GeodesicDistance7b} \\[6pt]
L_2 = R\log\left(\frac{(\textbf{y} - \textbf{x}_j)^2 + (y_3 - x_j)^2 + z_j^2}{z_j\epsilon}\right). \qquad \label{GeodesicDistance7c}
\end{IEEEeqnarray}
The ambient-channel correlation function of two local operators $O_1,O_2$ (scaling dimensions $\Delta_1,\Delta_2$) which are located at the boundary points $(\textbf{x}, x_3)$ and $(\textbf{y}, y_3)$ respectively, is given by the following formula \cite{KeranenSybesmaSzepietowskiThorlacius16, KimPalPark23}:
\begin{IEEEeqnarray}{c}
\langle O_1(\textbf{x},x_3)O_2(\textbf{y},y_3)\rangle = g_j\cdot\frac{e^{-L(\textbf{x},x_3;\textbf{y},y_3)}}{\epsilon^{\Delta_1+\Delta_2}}, \qquad \label{TwoPointFunctionGeodesic7}
\end{IEEEeqnarray}
where $g_j$ is the coupling constant at the cubic interaction point (dual to the junction point in the bulk). Writing $\Delta_j$ for the scaling dimension of a light exchange operator $O_j$, the total geodesic distance from the boundary to the interface brane in the bulk of AdS is
\begin{IEEEeqnarray}{c}
L(\textbf{x},x_3;\textbf{y},y_3) \equiv \frac{\Delta_j L_{jr}}{R} + \sum_{a=1}^2 \frac{\Delta_a L_a}{R}. \qquad \label{GeodesicDistance7d}
\end{IEEEeqnarray}
In the case of two identical operators ($\Delta_1 = \Delta_2 = \Delta$) which are located at the boundary points $(\textbf{x}, x_3)$ and $(-\textbf{x}, x_3)$, the geodesic length which is associated with their two-point function $\langle O(\textbf{x},x_3)O(-\textbf{x},x_3)\rangle$ becomes,
\begin{widetext}\begin{IEEEeqnarray}{ll}
L(\textbf{x},x_3;-\textbf{x},x_3) = 2\bigg[\Delta_j\log\left[\frac{\sqrt{(x_j - \kappa z_r)^2 + (z_j - z_r)^2} + \sqrt{(x_j - \kappa z_r)^2 + (z_j + z_r)^2}}{2\sqrt{z_j z_r}} \right] &+ \nonumber \\[6pt]
&\hspace{-1.5cm} + \Delta\log\left(\frac{\textbf{x}^2 + (x_j - x_3)^2 + z_j^2}{z_j\epsilon}\right)\bigg]. \qquad
\end{IEEEeqnarray}\end{widetext}
\noindent We may take $\textbf{x}_j = \textbf{x}_r = 0$ because the two-point function is invariant under $\textbf{x} \to -\textbf{x}$. To fix the junction and reflection points ($x_j,z_j,z_r$), we vary the geodesic length wrt $x_j$, $z_j$, and $z_r$. We find the following solution, which is valid for all values of $\xi \geq 0$ and $\kappa \geq 0$:
\begin{IEEEeqnarray}{c}
z_r = \sqrt{\frac{1 + \xi}{1 + \kappa^2}} \cdot x_3 , \quad x_j = x_3\sqrt{1+\xi}\cdot q, \nonumber \\[6pt]
z_j = x_3\sqrt{\left(1 + \xi\right)\left(1-q^2\right)},\qquad \ \label{JunctionReflection}
\end{IEEEeqnarray}
where $\xi = \textbf{x}^2/x_3^2 \geq 0$ and we have defined,
\begin{IEEEeqnarray}{c}
q \equiv \left|\frac{1 - r\sqrt{1 + \xi}}{\sqrt{1 + \xi} - r}\right|, \qquad r \equiv \frac{\Delta_j}{2\Delta}. \label{DefinitionQR}
\end{IEEEeqnarray}
To obtain the two-point function in the ambient channel we also assume,
\begin{IEEEeqnarray}{c}
2\Delta\sqrt{1 + \xi} \geq 2\Delta \geq \Delta_j\sqrt{1 + \xi} \geq \Delta_j,
\end{IEEEeqnarray}
or equivalently,
\begin{IEEEeqnarray}{c}
\sqrt{1 + \xi} \geq 1 \geq r\sqrt{1 + \xi} \geq r > 0.
\end{IEEEeqnarray}
This assumption allows us to remove the absolute value in \eqref{DefinitionQR} and the two-point function becomes,
\begin{IEEEeqnarray}{l}
\langle O(\textbf{x},x_3)O(-\textbf{x},x_3)\rangle = g_j\times\left(\frac{1 - r^2}{4\textbf{x}^2}\right)^{\Delta} \times \nonumber \\[6pt]
\hspace{1.5cm} \times \left(\frac{\sqrt{1 + \kappa^2} - \kappa q - \left|\kappa - q\sqrt{1 + \kappa^2}\right|}{\sqrt{1 - q^2}}\right)^{\Delta_j}. \qquad \label{TwoPointFunctionAmbient1}
\end{IEEEeqnarray}
For small values of $\xi$, the absolute value in \eqref{TwoPointFunctionAmbient1} should be taken with a minus overall sign so that the two-point function reads,
\begin{IEEEeqnarray}{l}
\langle O(\textbf{x},x_3)O(-\textbf{x},x_3)\rangle = g_j\times\left(\frac{1 - r^2}{4\textbf{x}^2}\right)^{\Delta} \times \nonumber \\[6pt]
\hspace{1cm} \times \left(\left(\kappa + \sqrt{1 + \kappa^2}\right)\cdot\sqrt{\frac{1-q}{1+q}}\right)^{\Delta_j}, \quad \xi \ll 1. \qquad \label{TwoPointFunctionAmbient2}
\end{IEEEeqnarray}
The perturbative expansion of the ambient two-point function which is obtained from the extremal point \eqref{JunctionReflection} becomes, in the limit of small $\xi$,
\begin{IEEEeqnarray}{c}
\langle O(\textbf{x},x_3)O(-\textbf{x},x_3)\rangle = \frac{1}{|2\textbf{x}|^{2\Delta}}\Bigg[1 + \sum_{j>0} (4\xi)^{\frac{\Delta_j}{2}} \CC_{\bullet\bullet}^j \times \nonumber \\[6pt]
\times \CC_j \left(1 - \frac{\Delta_j\xi}{4} + \frac{\Delta_j(\Delta_j+3)\xi^2}{32} - \ldots\right)\Bigg], \qquad
\end{IEEEeqnarray}
where $\Delta_{\{j=0\}} = 0$. This result fully agrees with the CFT prescription \eqref{TwoPointFunctionOPEheavy1} in the heavy operator limit $2\Delta > \Delta_j\gg1$ (and $r \gg 1$). We may also read off the holographic one and three-point function structure constants,
\begin{IEEEeqnarray}{c}
\CC_{\bullet\bullet}^j = g_j\left(1-r^2\right)^{\Delta}\left(\frac{1}{2}\sqrt{\frac{1 + r}{1 - r}}\right)^{\Delta_j}, \quad \CC_j = \left(\frac{c}{2}\right)^{\Delta_j}, \qquad \label{3ptStructureConstant}
\end{IEEEeqnarray}
where the constant $c$ is related to the one-point function structure constant $\CC_I$ which was defined in \eqref{DefinitionC}.
\subsubsection{Defect-channel two-point functions \label{SubSubSection:TwoPointDefect}}
\noindent In the defect channel, two ambient operators exchange boundary light modes, i.e.\ primary operators of the CFT on the codimension-1 boundary. See figure \ref{Figure4} below (valid as before for both the dCFT and the bulk of AdS).
\begin{figure}[H]\begin{center}\begin{tikzpicture}\begin{feynman}
\fill [lightgray] (3.3-.785,-1) rectangle (3.3+0.785,-1.2);
\vertex (a) [dot] at (3.3+.5,-.98) {};
\vertex (b) [empty dot] at (3.3-.5,.7) {};
\vertex (c) [empty dot] at (3.3+.5,.7) {};
\vertex (d) [dot] at (3.3-.5,-.98) {};
\vertex (f1) at (3.3-.9,-.98) {};
\vertex (f2) at (3.3+.9,-.98) {};
\vertex (O1) at (3.3-.5,1) {$O_1$};
\vertex (O2) at (3.3+.5,1) {$O_2$};
\vertex (Oj) at (3.3,-.6) {\color{Red}$\mathring{O}_j$};
\diagram* [line width=0.3mm, black] {(f1) -- (f2)};
\diagram* [line width=0.3mm, black] {(a) -- (c)};
\diagram* [line width=0.3mm, black] {(b) -- (d)};
\diagram* [line width=0.3mm, Red] {(a) -- (d)};
\end{feynman}\end{tikzpicture}\caption{Defect-channel two-point function.}\label{Figure4}\end{center}\end{figure}
\noindent The defect-channel two-point function can be computed by a recipe similar to \eqref{GeodesicDistance7a}--\eqref{GeodesicDistance7d}, which we used above to compute the two-point function in the ambient channel. This time, there is no bulk junction point but there are two reflection points on the interface brane. These are $(\textbf{x}_r, x_r, z_r)$ and $(\textbf{y}_r, y_r, u_r)$, with $x_r = \kappa\cdot z_r$ and $y_r = \kappa\cdot u_r$. For two identical operators ($\Delta_1 = \Delta_2 = \Delta$) which are placed at the AdS boundary points $(\textbf{x}, x_3)$ and $(\textbf{y}, y_3)$, the geodesic length which is associated with their defect-channel two-point function $\langle O(\textbf{x},x_3)O(\textbf{y},y_3)\rangle$ is 
\begin{widetext}\begin{IEEEeqnarray}{r}
L(\textbf{x},x_3;\textbf{y},y_3) = 2\rD_j\log\left[\frac{\sqrt{(\textbf{x}_r - \textbf{y}_r)^2 + (\kappa z_r - \kappa u_r)^2 + (z_r - u_r)^2} + \sqrt{(\textbf{x}_r - \textbf{y}_r)^2 + (\kappa z_r - \kappa u_r)^2 + (z_r + u_r)^2}}{2\sqrt{z_r u_r}} \right] + \nonumber \\[6pt]
+ \Delta\log\left(\frac{(\textbf{x} - \textbf{x}_r)^2 + (x_3 - \kappa z_r)^2 + z_r^2}{z_r\epsilon}\right) + \Delta\log\left(\frac{(\textbf{y} - \textbf{y}_r)^2 + (y_3 - \kappa u_r)^2 + u_r^2}{u_r\epsilon}\right), \qquad \label{GeodesicDistance8}
\end{IEEEeqnarray}\end{widetext}
where $\rD_j$ is the scaling dimension of the boundary operator $\mathring{O}_j$. The distance from the boundary of AdS$_5$ to the interface brane is minimized by the one-point function reflection points we computed in \eqref{ReflectionPoint1}:
\begin{IEEEeqnarray}{ll}
\textbf{x}_r = \textbf{x}, \quad &\quad z_r = \frac{x_3}{\sqrt{1 + \kappa^2}} > 0 \\
\textbf{y}_r = \textbf{y}, \quad &\quad u_r = \frac{y_3}{\sqrt{1 + \kappa^2}} > 0.
\end{IEEEeqnarray}
These correspond to two independent geodesic curves which leave the operators $O(\textbf{x},x_3)$ and $O(\textbf{y},y_3)$ at the boundary of AdS and terminate at the corresponding reflection points $(\textbf{x}_r, x_r = \kappa\cdot z_r, z_r)$ and $(\textbf{y}_r, y_r = \kappa\cdot u_r, u_r)$ on the interface brane. Plugging these reflection points into \eqref{GeodesicDistance8} and using the two-point function formula \eqref{TwoPointFunctionGeodesic7}, we find the following expression for the defect-channel two-point function:
\begin{IEEEeqnarray}{ll}
\langle O(\textbf{x},x_3)O(\textbf{y},y_3)\rangle = \frac{g_j^2\CC_I^2}{\left|x_3 y_3\right|^{\Delta}} \cdot \left[\frac{\sqrt{1 + \kappa^2}}{\sqrt{\tilde{\xi}} + \sqrt{1 + \tilde{\xi}}}\right]^{2\rD_j}, \qquad \label{TwoPointFunctionBoundary1}
\end{IEEEeqnarray}
where the definition of the one-point function structure constant $\CC_I$ can be found in \eqref{OnePointFunctionI2} and the modified invariant ratio $\tilde{\xi}$ is given by,
\begin{IEEEeqnarray}{l}
\tilde{\xi} \equiv (1 + \kappa^2)\cdot\frac{(x-y)^2}{4\left|x_3\right|\left|y_3\right|}.
\end{IEEEeqnarray}
Expanding the defect-channel two-point function \eqref{TwoPointFunctionBoundary1} for large values of the invariant ratio $\tilde{\xi} \gg 1$ we find,
\begin{IEEEeqnarray}{r}
\left\langle O\left(\textbf{x},x_3\right)O\left(\textbf{y},y_3\right)\right\rangle = \frac{1}{\left|x_3y_3\right|^{\rD}}\cdot\Bigg[\CC_I^2 + \sum_{j>0} \tilde{\xi}^{-\rD_j} \,\BB_{\bullet j}^2\times \nonumber \\[6pt]
\times \left(1 - \frac{\rD_j}{2\xi} +\frac{\rD_j(2\rD_j + 3)}{16\tilde{\xi}^2} - \ldots\right)\Bigg], \qquad \label{TwoPointFunctionBoundary2}
\end{IEEEeqnarray}
where $\rD_{\{j=0\}} = 0$, and the ambient-boundary coupling constant $\BB_{\bullet j}$ is given by,
\begin{IEEEeqnarray}{c}
\BB_{\bullet j}^2 = c\times\left(\frac{\sqrt{1 + \kappa^2}}{2}\right)^{\rD_j}, \quad c \equiv \frac{1}{\sqrt{1 + \kappa^2} - \kappa}, \qquad
\end{IEEEeqnarray}
while the definition of the constant $c$ was given in \eqref{DefinitionC} above. With these identifications, the perturbative expansion \eqref{TwoPointFunctionBoundary2} agrees with the generic expansion of the defect-channel two-point function \eqref{TwoPointFunctionOPEheavy2}, in the heavy operator limit $\tilde{\xi} \gg \rD_j \gg 1$. 
\section{Conclusions \label{Section:Conclusions}}
\noindent We studied holographic defect conformal field theories with bottom-up methods. We started by reviewing the computation of defect one-point functions in top-down holography, focusing on the D3-probe-D5 brane system. For this system, one-point functions of chiral primary operators (CPOs) have been computed at both weak and strong coupling and they have been found to agree with each other in a certain double-scaling limit. We then turned to the bottom-up approach. We showed that it leads to the same embedding of a codimension-1 interface brane in AdS space as the top-down method, after making the appropriate identifications. The bottom-up approach was subsequently applied to the computation of heavy scalar defect correlators at strong coupling, mainly one and two-point functions. In the former case, findings \cite{KastikainenShashi21} agree with the available top-down results at weak and strong coupling. In the latter case, the computed reflected, ambient and defect-channel two-point functions agree with the predictions of the operator product expansion (OPE) and the boundary operator expansion (BOE) in appropriate limits. Although ambient-channel two-point functions have been studied before with top-down methods (both at weak \cite{Widen17, deLeeuwIpsenKristjansenVardinghusWilhelm17} and strong coupling \cite{GeorgiouLinardopoulosZoakos23}), the study of reflected and defect-channel two-point functions with top-down methods (at both weak and strong coupling) is currently missing. \\
\indent There seem to be more solutions in the case of ambient-channel two-point functions which are worth exploring further. In the case of two identical operators ($\Delta_1 = \Delta_2 = \Delta$) which are placed at the boundary points $(\textbf{x}, x_3)$ and $(-\textbf{x}, x_3)$, there seems to be a solution to the minimization conditions for $\textbf{x}_j = \textbf{x}_r = 0$, and
\begin{IEEEeqnarray}{l}
z_r = \pm\sqrt{\frac{x_j^2 + z_j^2}{1 + \kappa^2}} \\ x_j = \frac{1}{3}\left(2x_3 \pm \sqrt{x_3^2 - 3\textbf{x}^2 + 3z_j^2}\right).
\end{IEEEeqnarray}
If instead, the two identical operators are placed at the boundary points $(\textbf{x}, x_3)$ and $(\textbf{x}, y_3)$, there seems to be another set of solutions for $\textbf{x}_j = \textbf{x}_r = \textbf{x}$, and
\begin{IEEEeqnarray}{l}
z_r = \pm\sqrt{\frac{x_j^2 + z_j^2}{1 + \kappa^2}}, \quad x_j = \pm\sqrt{x_3y_3 - z_j^2} \\
\text{or} \quad x_j = \frac{1}{3}\left(x_3 + y_3 \pm \sqrt{x_3^2 - x_3y_3 + y_3^2 + 3z_j^2}\right). \qquad
\end{IEEEeqnarray}
It would be interesting to find out whether the above solutions are legitimate by specifying the value of the junction coordinate $z_j$ which minimizes the total geodesic distance in \eqref{GeodesicDistance7a}--\eqref{GeodesicDistance7d}. If so, these solutions would give rise to new expressions for the two-point function in the ambient channel, which would be useful to determine. \\
\indent In the case of the defect-channel two-point function, we have chosen to explore only one solution which minimizes the distance of the interface brane to the boundary of AdS$_5$, by ignoring the distance between the two reflection points on the interface brane. It would be interesting to investigate whether more solutions exist, and minimize the total geodesic distance as well, i.e.\ by including the geodesic which has its endpoints on the interface brane. Indeed, there seems to be a solution when the two identical operators ($\Delta_1 = \Delta_2 = \Delta$) are placed at the boundary points $(\textbf{x}, x_3)$ and $(\textbf{x}, y_3)$, with $\textbf{x}_r = \textbf{y}_r = \textbf{x}$ and,
\begin{IEEEeqnarray}{l}
z_r = \frac{x_3 y_3}{u_r(1 + \kappa^2)}, \quad \text{or} \quad z_r = \frac{x_3(y_3 - \kappa u_r)}{\kappa y_3 - u_r\left(1 + \kappa^2\right)}. \qquad
\end{IEEEeqnarray}
It would be interesting to examine whether these solutions solve the corresponding minimization conditions (i.e.\ determine $u_r$) and if so, specify the corresponding value of the defect-channel two-point function. \\
\indent Although we have basically chosen to focus on the D3-D5 system, our results are completely general and most probably quite universal. We expect formulas very similar to the ones herein presented to hold in any codimension-1 holographic defect CFT, e.g.\ the dCFTs which are dual to the D3-D7, D3-NS5 and beta deformed D3-D5 probe brane system in four dimensions, or the D2-D4 system in three dimensions. It would be interesting to investigate whether the bottom-up approach can also be applied to holographic defect CFTs of higher codimensionalities, for example the codimension-2 dCFT which was recently proposed in \cite{GeorgiouLinardopoulosZoakos25} and is dual to a codimension-1 probe D3 brane in AdS space (D3-probe-D3 system). Other interesting systems of higher codimensionalities are the codimension-2 D3-D3 system, the codimension-3 D1-D3 system and other surface defects (such as for instance Gukov-Witten defects).\footnote{See also \cite{Rigatos25} for yet another interesting system where our present methods could be applied.} It should be possible to set up the computation of correlation functions in these systems with the bottom-up (geodesic approximation) approach. \\
\indent We expect a similar universality to hold for correlation functions. Although we have mainly examined heavy CPOs by computing their one and two-point correlation functions with the method of geodesic approximations, our results could apply to more complicated heavy operators. Another possible research direction would be to apply the method of geodesic approximations to the computation of spinorial correlation functions. Instead of computing correlation functions in AdS space, it would also be worth evaluating geodesic correlation functions in the compact counterparts of the geometry (e.g.\ the 5-sphere). See \cite{RodriguezGomez26} for a relevant recent work. Computing three and higher-point defect correlation functions with the methods of the present paper would also be very interesting. We hope to return to these issues in future projects.
\section{Acknowledgements}
\noindent We thank Hanse Kim for collaboration during the early stages of this work, Adolfo Holguin, Charlotte Kristjansen and Gordon Rogelberg for discussions. C.P.\ thanks APCTP for hospitality during a visit where part of this work was done. C.P.\ was supported by the National Research Foundation of Korea (NRF) grant funded by the Korea government (MSIT) [No.\ RS-2026-25473667]. G.L.\ was supported by the Research Start-up Fund of the Shanghai Institute for Mathematics and Interdisciplinary Sciences (SIMIS) [2302-SRFP-2026-0088]. The work of G.L.\ was also supported by the National Research Foundation of Korea (NRF) grant funded by the Korea government (MSIT) [No.\ 2023R1A2C1006975], as well as by an appointment to the JRG Program at the APCTP through the Science and Technology Promotion Fund and Lottery Fund of the Korean Government. G.L.\ thanks the participants of the joint program [APCTP-2025-J01] held at APCTP (Pohang, South Korea) for fruitful discussions.
\appendix\section{Chiral primary operators \label{AppendixA}}
\noindent The chiral primary operators (CPOs) of $\N = 4$ SYM are \quotes{light} $1/2$ BPS operators which are defined in terms of the six scalar fields $\phi_1,\ldots,\phi_6$ of $\N = 4$ SYM as follows:
\begin{IEEEeqnarray}{c}
O_{\II}(x) = \frac{(8\pi^2)^{\Delta/2}}{\lambda^{\Delta/2}\sqrt{\Delta}}\cdot C_{\II}^{i_1 i_2 \ldots i_{\Delta}} \tr\left[\phi_{i_1}(x) \, \ldots \, \phi_{i_{\Delta}}(x)\right], \qquad \label{CPOdefinition}
\end{IEEEeqnarray}
where $\II$ are $SO(6)$ quantum numbers which specify the CPO and $\Delta$ is the scaling dimension of the CPO. $C_{\II}^{i_1 i_2 \ldots i_{\Delta}}$ is a traceless symmetric tensor of $SO(6)$ which is normalized as $C_{\II}^{i_1 i_2 \ldots i_{\Delta}} C_{\JJ}^{i_1 i_2 \ldots i_{\Delta}} = \delta_{\II\JJ}$. The overall normalization of the CPOs \eqref{CPOdefinition} is such that the corresponding two-point function is normalized to unity \cite{LeeMinwallaRangamaniSeiberg98}:
\begin{IEEEeqnarray}{c}
\langle O_{\II}(x) O_{\JJ}(y)\rangle = \frac{\delta_{\II\JJ}}{|x-y|^{2\Delta}},
\end{IEEEeqnarray}
while the traceless symmetric tensor $C_{\II}^{i_1 i_2 \ldots i_{\Delta}}$ is associated with S$^5$ spherical harmonics,
\begin{IEEEeqnarray}{c}
C_{\II}^{i_1 i_2 \ldots i_{\Delta}} X_{i_1} X_{i_2} \ldots X_{i_{\Delta}} = Y_{\II}(X), \quad \sum_{i=1}^6 X_i^2 = 1. \qquad \label{SphericalHarmonicsSO6}
\end{IEEEeqnarray}
A very important operator among the CPOs \eqref{CPOdefinition} corresponds to the vacuum state of the $\N = 4$ SYM (ferromagnetic) spin chain which is defined as,
\begin{IEEEeqnarray}{c}
\mathcal{O} = \frac{(2\pi)^L}{\lambda^{L/2}\sqrt{L}}\cdot\text{tr}\left[Z^{L}\right], \quad Z \equiv \phi_1 + i\phi_2. \qquad \label{VacuumState}
\end{IEEEeqnarray}
It corresponds to a CPO of the form \eqref{CPOdefinition}--\eqref{SphericalHarmonicsSO6} with the corresponding spherical harmonic being given by,
\begin{IEEEeqnarray}{c}
Y_L = \left(\frac{X_1 + i X_2}{\sqrt{2}}\right)^L.
\end{IEEEeqnarray}
$Y_L$ is manifestly $SO(3)\times SO(3)$ symmetric under the following parametrization of the 5-sphere:
\begin{IEEEeqnarray}{ll}
X_1 = \cos\psi \cos\theta, \quad & X_2 = \sin\psi \cos\vartheta \qquad \label{SphereParametrization1} \\
X_3 = \cos\psi \sin\theta \cos\varphi, \quad & X_4 = \sin\psi \sin\vartheta \cos\chi \qquad \label{SphereParametrization2} \\
X_5 = \cos\psi \sin\theta \sin\varphi, \quad & X_6 = \sin\psi \sin\vartheta \sin\chi, \qquad \label{SphereParametrization3}
\end{IEEEeqnarray}
for which the metric of S$^5$ takes the form,
\begin{IEEEeqnarray}{c}
d\Omega_5^2 = d\psi^2 + \cos^2\psi \, d\Omega_2^2 + \sin^2\psi \, d\Omega_2^2. \label{S5metric}
\end{IEEEeqnarray}
Interestingly, both operators \eqref{CPOdefinition}, \eqref{VacuumState} can become arbitrarily heavy in the limit $L\sim\sqrt{\lambda}\sim\sqrt{N}$, in which case they are also known as BMN type of chiral primary operators (CPOs). \\
\indent The chiral primary operators \eqref{CPOdefinition}, \eqref{VacuumState} are very special among all operators of $\N=4$ SYM because they are protected from receiving quantum corrections in their scaling dimensions $\Delta$ (which are hence equal to their length $\Delta = L$), as well as their (two and) three-point function structure constants. This property ceases to hold for four and higher-point functions of CPOs \cite{EdenSchubertSokatchev00}. \\
\indent One-point functions of local operators, which trivially vanish in CFTs (and $\N = 4$ SYM), are no longer generically zero in the presence of codimension-1 defects. In the case of the CPOs \eqref{CPOdefinition}, \eqref{VacuumState}, the corresponding structure constants are also unprotected and do receive quantum corrections.
\section{Rescaled geometries \label{AppendixB}}
\noindent In this appendix we provide details on the derivation of solution \eqref{BraneEmbeddingBottomUp} with the bottom-up approach. We begin with a brief review of some basic concepts and definitions from the geometry of embedded hypersurfaces. \\
\indent It has been known for a long time that the consistent variational formulation of Einstein's field equations in spacetime manifolds with boundaries should include improvement boundary terms \cite{York72b, GibbonsHawking76}. The improved action of the gravitational field in a $d$-dimensional bulk spacetime $\MM$ (parametrized in terms of the coordinates $x_{\mu}$, $\mu = 0,\ldots, d-1$) with a codimension-1 non-null boundary $\partial\MM$ (parametrized in terms of the coordinates $\xi_a$, $a = 0, \ldots d-2$) reads:
\begin{IEEEeqnarray}{c}
S = S_{\text{EH}} + S_{\text{GHY}}, \label{ImprovedGravitationalAction1}
\end{IEEEeqnarray}
where we work in a system of units where $G = 1$, and set
\begin{IEEEeqnarray}{l}
S_{\text{EH}} = -\frac{1}{16\pi}\int_{\MM} d^d x \, \sqrt{-g} \left(R - 2\Lambda\right) \label{ImprovedGravitationalAction2} \\
S_{\text{GHY}} = - \frac{1}{8\pi}\int_{\partial\MM} d^{d-1}\xi \, \sqrt{-\gamma} \, K, \label{ImprovedGravitationalAction3}
\end{IEEEeqnarray}
where $g_{\mu\nu}$ is the bulk metric and $\Lambda$ the corresponding cosmological constant. In \eqref{ImprovedGravitationalAction3}, $\gamma$ stands for the determinant of the first fundamental form $\gamma_{\mu\nu}$ (to be defined in \eqref{FirstFundamentalFormDefinition} below), while $K = g^{\mu\nu} K_{\mu\nu} = \gamma^{\mu\nu} K_{\mu\nu}$ is the trace of the extrinsic curvature $K_{\mu\nu}$. The latter is defined as,
\begin{IEEEeqnarray}{c}
K_{\mu\nu} \equiv \gamma_{\mu}^{\rho}\gamma_{\nu}^{\sigma}\nabla_{\rho}\hat{n}_{\sigma} = \nabla_{\mu}\hat{n}_{\nu} - \hat{n}_{\mu}\hat{n}^{\rho}\nabla_{\rho}\hat{n}_{\nu},
\end{IEEEeqnarray}
where $\nabla_{\mu}$ is the covariant derivative wrt the bulk metric $g_{\mu\nu}$. By varying the improved gravitational action \eqref{ImprovedGravitationalAction1}--\eqref{ImprovedGravitationalAction3}, we obtain:
\small\begin{IEEEeqnarray}{ll}
\delta S = &-\frac{1}{16\pi}\int_{\MM} d^{d+1} x \sqrt{-g} \left\{R_{\mu\nu} - \frac{1}{2}\left(R - 2\Lambda\right)g_{\mu\nu}\right\}\delta g^{\mu\nu} - \nonumber \\[6pt]
& - \frac{1}{8\pi}\int_{\partial\MM} d^{d} \xi \cdot \bigg\{\frac{1}{2}\,\sqrt{-\gamma} \left(K_{\mu\nu} - K \gamma_{\mu\nu}\right)\delta \gamma^{\mu\nu} + \nonumber \\[6pt]
& \hspace{4.5cm} + \sqrt{-\gamma} \, \mathring{\nabla}_{\mu}\delta \hat{n}^{\mu}\bigg\}. \qquad \label{VariationGravitationalAction1}
\end{IEEEeqnarray}\normalsize
The bulk term gives rise to the Einstein's field equations which we assume are satisfied everywhere in the bulk $\MM$. The last term in \eqref{VariationGravitationalAction1} can be transformed into a boundary term on the $(d-2)$-dimensional non-null boundary $\partial^2\MM$ and is typically ignored. \\
\indent Supposing that the $(d-1)$-dimensional boundary is a domain wall, the total action of the system takes the form \eqref{ActionDW}:
\begin{IEEEeqnarray}{c}
S = S_{\text{EH}} + S_{\text{GHY}} + S_{\text{brane}}.
\end{IEEEeqnarray}
The corresponding variation becomes:
\small\begin{IEEEeqnarray}{ll}
\delta S = -\frac{1}{16\pi} \Bigg\{&\int_{\partial\MM_+} \sqrt{-\gamma^+} \, \pi_{\mu\nu}^+ \delta \gamma^{\mu\nu}_+ + \int_{\partial\MM_-} \sqrt{-\gamma^-} \, \pi_{\mu\nu}^-\delta \gamma^{\mu\nu}_- \nonumber \\
&- \int_{\partial\MM} \sqrt{-\gamma} \, t_{\mu\nu}\delta \gamma^{\mu\nu}\Bigg\}, \qquad \label{VariationGravitationalAction2}
\end{IEEEeqnarray}\normalsize
where the definitions of the energy-momentum tensor $t_{\mu\nu}$ and the canonical momentum $\pi_{\mu\nu}^{a}$, on either side of the domain wall ($a = \pm$), can be found in \eqref{EMT_Momentum} above. \\[6pt]
\indent We now consider an interface brane inside AdS$_5$ (i.e.\ $d = 4$; see \eqref{AdS5xS5metric} for the corresponding metric). The brane separates AdS$_5$ into two subregions AdS$_I$ and AdS$_{II}$, with different radii, $R_I, R_{II}$. The worldvolume coordinates of the interface brane are,
\begin{IEEEeqnarray}{c}
\xi_0 = x_0, \quad \xi_1 = x_1, \quad \xi_2 = x_2, \quad \xi_3 = x_3, \label{BraneWorldvolume}
\end{IEEEeqnarray}
while the ansatz for the embedding of the interface brane inside AdS$_5$ is identical with the top-down approach \cite{KarchRandall01b}:
\begin{IEEEeqnarray}{c}
z = z(x_3). \label{BraneEmbedding2}
\end{IEEEeqnarray}
The unit normal vector $\hat{n}_{\mu}$ is found to be ($\acute{z} \equiv \partial_3 z$),
\begin{IEEEeqnarray}{c}
\hat{n}_{\mu}^{\pm} = \frac{\pm R_{\pm}}{z\sqrt{1+\acute{z}^2}}\cdot\left(0,0,0,-\acute{z},1\right),
\end{IEEEeqnarray}
where $R_{\pm}$ is any of the two AdS$_5$ radii, $R_+ = R_I$ and $R_- = R_{II}$. The first fundamental form $\gamma_{\mu\nu}$ and the normal projector $N_{\mu\nu}$ are defined as,
\begin{IEEEeqnarray}{c}
\gamma_{\mu\nu} = g_{\mu\nu} - \hat{n}_{\mu}\hat{n}_{\nu}, \qquad N_{\mu\nu} = \hat{n}_{\mu}\hat{n}_{\nu}, \label{FirstFundamentalFormDefinition}
\end{IEEEeqnarray}
which leads to the following expression for the first fundamental form of the interface brane \eqref{BraneWorldvolume}--\eqref{BraneEmbedding2}:
\begin{IEEEeqnarray}{ll}
\gamma_{\mu\nu}^{\pm} = \frac{R_{\pm}^2}{z^2}\cdot\bigg\{&\text{diag}\left(-1,1,1,0,0\right) + \nonumber \\[6pt]
& + \frac{1}{1+\acute{z}^2}\cdot\text{block}\big(0_{3\times3},\left[\begin{array}{cc}1&\acute{z}\\\acute{z}&\acute{z}^2\end{array}\right]\big)\bigg\}, \qquad
\end{IEEEeqnarray}
while the corresponding momenta $\pi^{\pm}_{\mu\nu}$ are given by,
\begin{IEEEeqnarray}{l}
\pi^{\pm}_{\mu\nu} = \frac{3}{R_{\pm}\sqrt{1 + \acute{z}^2}}\cdot\gamma^{\pm}_{\mu\nu} + \nonumber \\
\hspace{2cm} + \frac{R_{\pm} \aacute{z}}{z\left(1 + \acute{z}^2\right)^{3/2}}\cdot\text{diag}\left(-1,1,1,0,0\right), \qquad \label{CanonicalMomenta}
\end{IEEEeqnarray}
The first fundamental forms on either side of the domain wall are therefore related as follows:
\begin{IEEEeqnarray}{c}
\gamma_{\mu\nu}^- = \frac{R^2_-}{R^2_+}\cdot \gamma_{\mu\nu}^+, \qquad \sqrt{-\gamma^-} = \frac{R^5_-}{R^5_+}\cdot\sqrt{-\gamma^+}. \qquad
\end{IEEEeqnarray}
\indent At this point, we also assume that the energy-momentum tensor of the interface brane is simply related to its first fundamental form \eqref{EMT_Ansatz}:
\begin{IEEEeqnarray}{c}
t_{\mu\nu} = \sigma\cdot\gamma_{\mu\nu},
\end{IEEEeqnarray}
where $\sigma$ is the tension of the brane. Therefore the energy-momentum tensors on the two sides of the domain wall are also related as
\begin{IEEEeqnarray}{c}
t_{\mu\nu}^- = \sigma \cdot \gamma_{\mu\nu}^- = \sigma \cdot \frac{R^2_-}{R^2_+} \, \gamma_{\mu\nu}^+ = \frac{R^2_-}{R^2_+}\cdot t_{\mu\nu}^+. \qquad
\end{IEEEeqnarray}
\indent Assuming that the variations of the first fundamental form are the same on the two sides of the boundary, i.e.\ $\delta \gamma^{\mu\nu}_+ = \delta \gamma^{\mu\nu}_- = \delta \gamma^{\mu\nu}$, the variation of the improved gravitational action \eqref{VariationGravitationalAction2} becomes:
\begin{IEEEeqnarray}{ll}
\delta S = - \frac{1}{16\pi} \int_{\partial\MM_+} &\sqrt{-\gamma^+}\cdot\Bigg[\pi_{\mu\nu}^+ - \frac{R^5_-}{R^5_+} \, \pi_{\mu\nu}^- - \nonumber \\[6pt]
& \hspace{.5cm} - \left(1 + \frac{R^7_-}{R^7_+}\right) t_{\mu\nu}^+\Bigg]\cdot\delta \gamma^{\mu\nu}_+. \qquad \label{VariationGravitationalAction3}
\end{IEEEeqnarray}
Note that the resulting configuration is equivalent to one where all matter is concentrated on one side of the boundary. The variation \eqref{VariationGravitationalAction3} implies the following matching condition:
\begin{IEEEeqnarray}{ll}
R^5_+ \, \pi_{\mu\nu}^+ - R^5_- \, \pi_{\mu\nu}^- = \left(R^5_+ + \frac{R^7_-}{R^2_+}\right) \, \sigma\cdot \gamma_{\mu\nu}^+, \label{MatchingCondition1}
\end{IEEEeqnarray}
which can also be written in the form \eqref{IsraelMatchingConditions}:
\begin{IEEEeqnarray}{ll}
\hat{\pi}_{\mu\nu}^I - \hat{\pi}_{\mu\nu}^{II} = t_{\mu\nu}, \qquad \label{MatchingCondition2}
\end{IEEEeqnarray}
where we have defined,
\begin{IEEEeqnarray}{ll}
\hat{\pi}_{\mu\nu}^{I/II} \equiv R^5_{\pm} \pi_{\mu\nu}^{\pm}, \quad t_{\mu\nu} \equiv \left(R^5_+ + \frac{R^7_-}{R^2_+}\right) \, \sigma\cdot \gamma_{\mu\nu}^+. \qquad \label{EMT_Redefinition}
\end{IEEEeqnarray}
\indent To solve the matching condition \eqref{MatchingCondition2}, we plug the canonical momenta \eqref{CanonicalMomenta} and the energy-momentum tensor \eqref{EMT_Redefinition} into the matching condition equations \eqref{MatchingCondition1}--\eqref{MatchingCondition2}. To solve, we obviously have to set $\aacute{z} = 0$ in \eqref{CanonicalMomenta} which leads to,
\begin{IEEEeqnarray}{c}
\frac{3R^4_+}{\sqrt{1 + \acute{z}^2}}\cdot\gamma_{\mu\nu}^{+} - \frac{3R^6_-}{R^2_{+}\sqrt{1 + \acute{z}^2}}\cdot\gamma_{\mu\nu}^+ = t_{\mu\nu}. \qquad \label{JunctionEquation2}
\end{IEEEeqnarray}
With a few redefinitions, \eqref{JunctionEquation2} can be brought in the form \eqref{JunctionEquation1}:
\begin{IEEEeqnarray}{l}
\frac{3}{\sqrt{1 + \acute{z}^2}}\cdot\left(\frac{1}{r_+} - \frac{1}{r_-}\right) = \sigma, \quad r_{\pm} \equiv R_{\pm} + \frac{R_{\mp}^7}{R_{\pm}^6}, \qquad \label{JunctionEquation3}
\end{IEEEeqnarray}
which we can solve, obtaining the solution \eqref{BraneEmbeddingBottomUp}--\eqref{BraneTension}:
\begin{IEEEeqnarray}{c}
x_3 = \kappa\cdot z, \qquad \kappa = \frac{\sigma \, r_+ r_-}{\sqrt{9\left(r_+ - r_-\right)^2 - \sigma^2r_+^2 r_-^2}}. \qquad
\end{IEEEeqnarray}
\section{Crossing equation \label{AppendixC}}
\noindent The generic form of two-point functions in codimension-1 defect CFTs is \cite{McAvityOsborn95}:
\begin{IEEEeqnarray}{l}
\left\langle O_1\left(\textbf{x},x_3\right)O_2\left(\textbf{y},y_3\right)\right\rangle = \frac{f_{12}\left(\xi\right)}{\left|x_3\right|^{\Delta_1}\left|y_3\right|^{\Delta_2}}, \qquad \label{TwoPointFunction}
\end{IEEEeqnarray}
where the codimension-1 invariant ratio is defined as,
\begin{IEEEeqnarray}{l}
\xi \equiv \frac{(x-y)^2}{4\left|x_3\right|\left|y_3\right|}.
\end{IEEEeqnarray}
The function $f_{12}(\xi)$ which shows up as two-point function structure constant can be completely specified in the so-called ambient channel by using the ambient OPE (see e.g.\ \cite{Linardopoulos25a} for details). In four dimensions,
\begin{IEEEeqnarray}{ll}
f_{12}\left(\xi\right) =& \left(4\xi\right)^{-\frac{\Delta_1 + \Delta_2}{2}} \Bigg[\delta_{12} + \sum_j (4\xi)^{\frac{\Delta_j}{2}}\,\CC_{12}^j \, \CC_j \times \nonumber \\[6pt]
& \times{_2}F_1\Big(\frac{\Delta_j + \delta\Delta}{2}, \frac{\Delta_j - \delta\Delta}{2},\Delta_j - 1;-\xi\Big)\Bigg], \qquad \label{ConformalBlockAmbientChannel}
\end{IEEEeqnarray}
where $\delta\Delta \equiv \Delta_1-\Delta_2$ and the sum is over all CFT primary operators $O_j$, whose dCFT one-point function structure constant is $\CC_j$, and their CFT three-point function structure constant corresponding to $\langle O_1 O_2 O_j\rangle$ is $\CC_{12}^j$. For operators with equal scaling dimensions $\Delta_1 = \Delta_2 = \Delta$ (and $\CC_{12}^j = \CC_{\bullet\bullet}^j$) the two-point function reads,
\begin{IEEEeqnarray}{ll}
\left\langle O\left(\textbf{x},x_3\right)O\left(\textbf{y},y_3\right)\right\rangle = &\frac{1}{(x-y)^{2\Delta}}\,\Bigg[1 + \sum_j (4\xi)^{\frac{\Delta_j}{2}} \, \CC_{\bullet\bullet}^j \, \CC_j \times \nonumber \\[6pt]
&\times{_2}F_1\Big(\frac{\Delta_j}{2}, \frac{\Delta_j}{2},\Delta_j - 1;-\xi\Big)\Bigg]. \qquad \label{TwoPointFunctionOPE}
\end{IEEEeqnarray}
We may expand the ambient-channel two-point function in small values of $\xi \ll 1$ finding,
\begin{IEEEeqnarray}{c}
\left\langle O\left(\textbf{x},x_3\right)O\left(\textbf{y},y_3\right)\right\rangle = \frac{1}{(x-y)^{2\Delta}}\Bigg[1 + \sum_j (4\xi)^{\frac{\Delta_j}{2}} \, \CC_{\bullet\bullet}^j \, \CC_j\times \nonumber\\[6pt]
\times\left(1 - \frac{\Delta_j^2\xi}{4(\Delta_j-1)} + \frac{\Delta_j(\Delta_j+2)^2\xi^2}{32(\Delta_j-1)} - \ldots\right)\Bigg]. \qquad 
\end{IEEEeqnarray}
If we further assume that the operators $O_j$ are heavy, i.e.\ $\Delta_j \gg 1$, the two-point function becomes, in the limit of small $\Delta_j\xi \ll 1$:
\begin{IEEEeqnarray}{ll}
\left\langle O\left(\textbf{x},x_3\right)O\left(\textbf{y},y_3\right)\right\rangle = &\frac{1}{(x-y)^{2\Delta}}\Bigg[1 + \sum_j (4\xi)^{\frac{\Delta_j}{2}} \, \CC_{\bullet\bullet}^j \, \CC_j \times \nonumber\\[6pt]
&\hspace{-1.5cm}\times\underbrace{\left(1 - \frac{\Delta_j\xi}{4} + \frac{\Delta_j^2\xi^2}{32} - \frac{\Delta_j^3 \xi^3}{384} + \ldots\right)}_{\exp(-\Delta_j\xi/4)}\Bigg]. \qquad \label{TwoPointFunctionOPEheavy1}
\end{IEEEeqnarray}
\indent Alternatively, the two-point function structure constant $f_{12}(\xi)$ can be computed in the so-called defect channel by means of the boundary operator expansion (BOE). In four dimensions,
\begin{IEEEeqnarray}{ll}
f_{12}\left(\xi\right) = &\CC_1\CC_2 + \sum_j \xi^{-\rD_j}\,\BB_{1j}\BB_2^j \times \nonumber \\[6pt]
& \times{_2}F_1\Big(\rD_j, \rD_j - 1, 2\rD_j - 2;-\xi^{-1}\Big), \qquad\label{ConformalBlockDefectChannel}
\end{IEEEeqnarray}
where $\CC_1,\CC_2$ are the one-point function structure constants of the ambient operators $O_1, O_2$, and the sum is over all boundary primaries $\mathring{O}_j$ whose scaling dimension is $\rD_j$. $\BB_{1j}$ and $\BB_{2j}$ are the structure constants of the ambient-boundary two-point functions $\langle O_1 \mathring{O}_j\rangle$ and $\langle O_2 \mathring{O}_j\rangle$. For $i = 1,2$ we also define $\BB_i^j \equiv \BB_{ij}/\mathring{\CC}_{jj}$, where $\mathring{\CC}_{jj}$ is the structure constant of the two-point function $\langle\mathring{O}_j\mathring{O}_j\rangle$.
\begin{figure}[H]\begin{center}\begin{tikzpicture}\begin{feynman}
\fill [lightgray] (-.785,-1) rectangle (+0.785,-1.2);
\vertex (a) [dot] at (0,0) {};
\vertex (b) [empty dot] at (-.5,.7) {};
\vertex (c) [empty dot] at (+.5,.7) {};
\vertex (d) [dot] at (0,-.98) {};
\vertex (f1) at (-.9,-.98) {};
\vertex (f2) at (+.9,-.98) {};
\vertex (O1) at (-.7,1) {$O_1$};
\vertex (O2) at (+.7,1) {$O_2$};
\vertex (Oj) at (+.3,-.5) {\color{Blue}$O_j$};
\vertex (sum) at (-1.4,-.5) {\huge$\sum\limits_{\scaleto{j\mathstrut}{10pt}}$};
\vertex (correlator) at (-2.7,-.25) {$\left\langle O_1O_2\right\rangle = $};
\vertex (equals) at (1.4,-.3) {$=$};
\diagram* [line width=0.3mm, Red] {(c) -- (a) -- (b)};
\diagram* [line width=0.3mm, Blue] {(a) -- (d)};
\diagram* [line width=0.3mm, black] {(f1) -- (f2)};
\fill [lightgray] (3.3-.785,-1) rectangle (3.3+0.785,-1.2);
\vertex (a) [dot] at (3.3+.5,-.98) {};
\vertex (b) [empty dot] at (3.3-.5,.7) {};
\vertex (c) [empty dot] at (3.3+.5,.7) {};
\vertex (d) [dot] at (3.3-.5,-.98) {};
\vertex (f1) at (3.3-.9,-.98) {};
\vertex (f2) at (3.3+.9,-.98) {};
\vertex (O1) at (3.3-.5,1) {$O_1$};
\vertex (O2) at (3.3+.5,1) {$O_2$};
\vertex (Oj) at (3.3,-.6) {\color{Red}$\mathring{O}_j$};
\vertex (sum) at (3-1,-.5) {\huge$\sum\limits_{\scaleto{j\mathstrut}{10pt}}$};
\diagram* [line width=0.3mm, black] {(f1) -- (f2)};
\diagram* [line width=0.3mm, black] {(a) -- (c)};
\diagram* [line width=0.3mm, black] {(b) -- (d)};
\diagram* [line width=0.3mm, Red] {(a) -- (d)};
\end{feynman}\end{tikzpicture}\caption{Crossing equation in codimension-1 dCFT.}\label{Figure5}\end{center}\end{figure}
\noindent Equality between the ambient-channel two-point function structure constant \eqref{ConformalBlockAmbientChannel} and the defect-channel one \eqref{ConformalBlockDefectChannel} leads to the so-called crossing equation (see figure \ref{Figure5}). The corresponding defect-channel expression for the two-point function \eqref{TwoPointFunction} in case the two operators are the same ($\Delta_1 = \Delta_2 = \Delta$, $\CC_1 = \CC_2 = \CC$) is,
\begin{IEEEeqnarray}{c}
\left\langle O\left(\textbf{x},x_3\right)O\left(\textbf{y},y_3\right)\right\rangle = \frac{1}{\left|x_3y_3\right|^{\Delta}}\cdot\Bigg[\CC^2 + \sum_j \xi^{-\rD_j}\,\BB_{\bullet j}^2 \times \nonumber \\[6pt]
\hspace{2cm} \times {_2}F_1\Big(\rD_j, \rD_j - 1, 2\rD_j - 2;-\xi^{-1}\Big)\Bigg]. \qquad \label{TwoPointFunctionBOE}
\end{IEEEeqnarray}
Expanding the defect-channel two-point function for large values of $\xi \gg 1$, we find,
\begin{IEEEeqnarray}{r}
\left\langle O\left(\textbf{x},x_3\right)O\left(\textbf{y},y_3\right)\right\rangle = \frac{1}{\left|x_3y_3\right|^{\rD}}\cdot\Bigg[\CC^2 + \sum_j \xi^{-\rD_j} \,\BB_{\bullet j}^2\times \nonumber \\[6pt]
\times \left(1 - \frac{\rD_j}{2\xi} + \frac{(\rD_j+1)\rD_j^2}{4(2\rD_j-1)\xi^2} - \ldots\right)\Bigg]. \qquad
\end{IEEEeqnarray}
For heavy boundary operators $\mathring{O}_j$, the two-point function becomes, in the limit $\xi \gg \rD_j \gg 1$:
\begin{IEEEeqnarray}{r}
\left\langle O\left(\textbf{x},x_3\right)O\left(\textbf{y},y_3\right)\right\rangle = \frac{1}{\left|x_3y_3\right|^{\rD}}\cdot\Bigg[\CC^2 + \sum_j \xi^{-\rD_j} \,\BB_{\bullet j}^2\times \nonumber \\[6pt]
\times \underbrace{\left(1 - \frac{\rD_j}{2\xi} + \frac{\rD_j^2}{8\xi^2} - \frac{\rD_j^2}{48\xi^3} + \ldots\right)}_{\exp(-\rD_j/2\xi)}\Bigg], \qquad \quad \label{TwoPointFunctionOPEheavy2}
\end{IEEEeqnarray}
taking the characteristic exponential form of heavy operator expansions, c.f.\ \eqref{OnePointFunctionCPOstrong2}, \eqref{OnePointFunctionCPOstrong3}, and \eqref{TwoPointFunctionOPEheavy1}.

\end{document}